\documentclass[aps,prb,reprint,superscriptaddress,floatfix]{revtex4-2}

\usepackage{amsmath,amssymb, bbm, dsfont}
\usepackage{graphicx}
\usepackage[nopatch=footnote]{microtype}
\usepackage{hyperref}
\usepackage{xcolor}
\hypersetup{colorlinks=true,linkcolor=blue,urlcolor=blue,citecolor=red}

\newcommand{\average}[1]{\langle #1\rangle}

\newcommand{\dket}[1]{|#1\rangle\!\rangle}

\makeatletter
\newif\ifinsupplementaltoc
\newcommand{\printsupplementaltableofcontents}{%
	\begingroup
	\setcounter{tocdepth}{2}%
	\insupplementaltocfalse
	\let\supplemental@contentsline\contentsline
	\def\supplemental@title{\numberline {}Supplemental Material}%
	\def\supplemental@contents{\numberline {}Contents}%
	\def\contentsline##1##2##3##4{%
		\def\supplemental@entry{##2}%
		\ifinsupplementaltoc
			\ifx\supplemental@entry\supplemental@contents
			\else
				\supplemental@contentsline{##1}{##2}{##3}{##4}%
			\fi
		\else
			\ifx\supplemental@entry\supplemental@title
				\global\insupplementaltoctrue
			\fi
		\fi
	}%
	\tableofcontents
	\endgroup
}
\makeatother

\begin{document}

\title{Two-parameter Family--Vicsek scaling in a dissipative XXZ spin chain}

\author{C\u{a}t\u{a}lin Pa\c{s}cu Moca}
\email{mocap@uoradea.ro}
\affiliation{Department of Physics, University of Oradea, 410087, Oradea, Romania}
\affiliation{Department of Theoretical Physics, Institute of Physics, Budapest University of Technology and Economics, M\H{u}egyetem rkp.~3, H-1111 Budapest, Hungary}
\affiliation{MTA-BME Lend\"ulet ``Momentum'' Open Quantum Systems Research Group, Institute of Physics, Budapest University of Technology and Economics, M\H{u}egyetem rkp.~3, H-1111 Budapest, Hungary}

\author{Doru Sticlet}
\affiliation{National Institute for R\&D of Isotopic and Molecular Technologies, 67-103 Donat, 400293 Cluj-Napoca, Romania}

\author{Tam\'as Vicsek}
\affiliation{Department of Biological Physics, Institute of Physics, E\"otv\"os University, Budapest, H-1117, Hungary}

\author{Bal\'azs D\'ora}
\affiliation{Department of Theoretical Physics, Institute of Physics, Budapest University of Technology and Economics, M\H{u}egyetem rkp.~3, H-1111 Budapest, Hungary}
\affiliation{MTA-BME Lend\"ulet ``Momentum'' Open Quantum Systems Research Group, Institute of Physics, Budapest University of Technology and Economics, M\H{u}egyetem rkp.~3, H-1111 Budapest, Hungary}

\date{\today}

\begin{abstract}
Family--Vicsek (FV) scaling provides an understanding for the growth and finite-size saturation of fluctuations in classical systems. 
Here, we extend the FV roughness to transferred segment magnetization after quantum quenches in a dissipative XXZ spin chain with homogeneous gain and loss, starting from 
a nonequilibrium steady state with finite magnetization. 
In the non-interacting limit, we derive a closed-form expression for the roughness in the presence of  dissipation.
It displays two-parameter FV scaling and smoothly interpolates between the clean ballistic behavior and the dissipation dominated scalings.
For interacting chains, tensor-network simulations show that the non-dissipative  ballistic growth at finite magnetization is robust, 
whereas the full Lindblad evolution is generically controlled by the dissipative relaxation time and exhibits a dissipation-dominated collapse.
\end{abstract}

\maketitle

\paragraph*{Introduction.\textemdash}
Understanding universal features of far-from-equilibrium dynamics is a central theme in many-body physics~\cite{nishiura2002far,Petruccione.2002}.
In isolated quantum systems, a particularly clean route is a quantum quench~\cite{Mitra.2018}, which triggers nontrivial spatio-temporal 
spreading of correlations and fluctuations. 
An interesting question is whether this quantum dynamics can be organized by scaling laws similar to those that govern 
classical nonequilibrium phenomena.

The Family--Vicsek (FV) scaling ansatz, introduced previously by one of 
the present authors in the context of classical surface growth~\cite{s1992fractal,family1985scaling}, captures how the roughness of a fluctuating quantity in a subsystem of linear size $\ell$ grows in time and ultimately 
saturates due to finite-size effects~\cite{barabasi1995fractal}. In quantum spin chains, an analogous roughness can be defined from fluctuations of an extensive observable 
within the segment, e.g., the segment magnetization $\Sigma_\ell$ or, equivalently, the transferred segment magnetization $\Delta\Sigma_\ell(t)
\equiv\Sigma_\ell(t)-\Sigma_\ell(0)$. Denoting the resulting roughness by $W(\ell,t)$, FV scaling indicates that $W(\ell,t)$ can be expressed in 
terms of a single scaling variable $t/\ell^{z}$ as
\begin{equation}
   W(\ell,t) \sim \ell^{\alpha}\, f\!\left(\frac{t}{\ell^{z}}\right),\label{eq:FV_scaling}
\end{equation}
with $f(x\ll 1)\sim x^{\beta}$ and $f(x\gg 1)\to\mathrm{const}$. The exponent $\alpha$ controls the subsystem-size dependence of the saturated roughness, while $\beta$ governs the early-time growth, and $z=\alpha/\beta$ is the dynamical exponent.

Recently, Ref.~\cite{kwon.2026} demonstrated FV scaling experimentally 
in quantum gases far from equilibrium, establishing FV collapses in one-dimensional quantum dynamics and emphasizing their connection to transport universality classes, 
ranging from ballistic spreading to diffusion and KPZ-type superdiffusion~\cite{KPZ.86,Nahum2017,fujimoto2020family,fujimoto2021dynamical,
glidden2021bidirectional,fujimoto2022impact,aditya2024family,bhakuni2024dynamic,cecile2024squeezed,Tater2025,Minoguchi2025,Moca.2026}. 
This connection can depend sensitively on integrability, conservation laws, and symmetries, and it can be modified by noise and decoherence~\cite{Bua.2012,Cai2013,Essler2016,Medvedyeva2016,Bertini.2021,Landi2022,Fazio2025}.

Motivated by these developments, we study FV scaling in an open quantum many-body setting where coherent XXZ dynamics competes with bulk 
gain/loss described by a Markovian Lindblad equation~\cite{Gorini.1976,Lindblad.1976}. The drive prepares an analytically tractable, 
product-form nonequilibrium steady state (NESS) that is featureless in equal-time connected correlations, yet provides a natural 
high-entropy reference state for quench dynamics. Because gain/loss breaks conservation of total magnetization, the open-system dynamics can 
differ qualitatively from the closed-chain transport classification, suggesting a crossover toward Edwards--Wilkinson-type scaling~\cite{edwards1982surface}.

Our main result is the construction of a two-parameter FV scaling 
in the presence of dissipation, with  the roughness expressed as
\begin{equation}
   W(\ell,t) \sim \ell^{\alpha}\, f\!\left(\frac{t}{\ell^{z}},\frac{t}{t_\Gamma}\right),\label{eq:2FV_scaling}
\end{equation}
where $t_\Gamma$ represents a dissipative timescale.
We obtain an explicit expression for the roughness at finite
dissipation rate in the non-interacting limit, and show that it reduces in the unitary limit (=no dissipation) to ballistic FV scaling with $\alpha=\beta=1/2$ and
$z=1$, preceded by a short-time transient with the roughness scaling as $\propto t$.
For the full Lindblad evolution in the non-interacting limit, the coherent structure survives but correlations are exponentially damped, 
leading to a saturation controlled by $t_\Gamma$.
With interactions, FV scaling under unitary evolution remains robust: the scaling  collapse of initially magnetized systems is compatible with
ballistic dynamics, while for non-magnetized initial state,  the dynamical exponent crosses over from ballistic through KPZ-type to
diffusive behavior~\cite{Moca.2026} with increasing interaction. Upon breaking integrability with a next-nearest-neighbor coupling, the unitary ballistic
scenario persists for magnetized initial states, whereas the non-magnetized initial state with  integrability breaking yields diffusion over a
broad range of interactions. By contrast, for interacting Lindbladian dynamics the one-parameter FV collapse in the coherent
scaling variable breaks down and the observed collapse is governed primarily by $\Gamma t$.

\paragraph*{Model.\textemdash}
We consider Markovian evolution $\partial_t\rho=\mathcal{L}[\rho]\equiv-i[H,\rho]+\mathcal{D}[\rho]$~\cite{Gorini.1976,Lindblad.1976}, with XXZ Hamiltonian
\begin{equation}
H=\frac{J}{4}\sum_l\big(\sigma_l^x\sigma_{l+1}^x+\sigma_l^y\sigma_{l+1}^y+\Delta\,\sigma_l^z\sigma_{l+1}^z\big),
\label{eq:H_xxz}
\end{equation}
and homogeneous jump operators $L_l^-=\sqrt{\gamma_l}\,\sigma_l^-$, $L_l^+=\sqrt{\gamma_p}\,\sigma_l^+$, with loss (gain) rates $\gamma_l$ 
($\gamma_p$). The dissipator has the standard Lindblad form $\mathcal{D}[\rho]=\sum_{l,\eta=\pm}\big(L_l^{\eta}\rho L_l^{\eta\dagger}-\tfrac12\{L_l^{\eta\dagger}L_l^{\eta},\rho\}\big)$. For $\gamma_l,\gamma_p>0$ the dynamics relaxes to a unique nonequilibrium steady state 
(NESS) that is remarkably simple: it is diagonal in the $\sigma^z$ basis, independent of $J$ and $\Delta$, and factorizes as $\rho_{\mathrm{SS}}=\bigotimes_l\rho_l$ with $\rho_l=(1+\zeta)^{-1}\,\mathrm{diag}(\zeta,1)$, where $\zeta\equiv\gamma_p/\gamma_l$ plays the role of a 
spin fugacity. The steady state is therefore an infinite-temperature grand-canonical ensemble with local magnetization $\langle\sigma_l^z\rangle_{\mathrm{SS}}=(\zeta-1)/(\zeta+1)$; for $\zeta=1$ one has $\rho_{\mathrm{SS}}\propto \mathds{1}$ and vanishing average magnetization. 
Performing the Jordan--Wigner transformation~\cite{Jordan.1928}, the model can be viewed as a chain of spinless fermions with 
uniform filling $\bar n\equiv\langle n_l\rangle_{\mathrm{SS}}=\zeta/(1+\zeta)$ . While the Hamiltonian has a $\mathrm{U}(1)$ 
symmetry generated by $S^z_{\mathrm{tot}}=\tfrac12\sum_l\sigma_l^z$, bulk gain/loss breaks magnetization conservation (and thus the corresponding hydrodynamics), except for the special spin-rotation symmetry at $\zeta=1$. 
Finally, any additional coherent term that conserves $S^z_{\mathrm{tot}}$ (e.g., a next-nearest-neighbor coupling) leaves the NESS unchanged and affects only the transient post-quench 
dynamics. The overall dissipative scale is set by the total rate $\Gamma\equiv\gamma_l+\gamma_p$, which defines the microscopic relaxation time $t_\Gamma\sim\Gamma^{-1}$.
 (further details are given in the Supplemental Material (SM)~\cite{SM}).

\paragraph*{Vectorized approach.\textemdash}
We probe fluctuations via a two-time quantum generating function (QGF)~\cite{Moca.2026,Valli.2025}, closely related to work statistics and full counting statistics~\cite{Talkner2007,Silva.2008,Levitov1993,Klich2009}. For a segment of length $\ell$ we define $\Sigma_\ell\equiv\tfrac12\sum_{j\in\mathrm{seg}(\ell)}\sigma_j^z$ and the transferred magnetization $\Delta\Sigma_\ell(t)\equiv\Sigma_\ell(t)-\Sigma_\ell(0)$. The QGF is
\begin{equation}
G_\ell(\lambda,t)=\mathrm{Tr}\!\left[e^{i\lambda\Sigma_\ell}\,e^{\mathcal{L}t}\!\left(e^{-i\lambda\Sigma_\ell}\rho_{\mathrm{SS}}\right)\right],
\label{eq:QGF}
\end{equation}
and cumulants follow from $\kappa_n=( -i)^n\,\partial_\lambda^n\ln G_\ell\vert_{\lambda=0}$, with roughness $W(\ell,t)\equiv\sqrt{\kappa_2(\ell,t)}$. Because 
$\Delta\Sigma_\ell(t)$ is a difference of operators at times $0$ and $t$, $G_\ell(\lambda,t)$ is intrinsically a two-time object ($\sim \langle e^
{i\lambda\Sigma_\ell(t)}e^{-i\lambda\Sigma_\ell(0)}\rangle$) and therefore should be viewed as a characteristic function defined by a counting-field protocol 
rather than as the expectation value of a single-time Hermitian observable~\cite{Talkner2007,Silva.2008,Levitov1993,Klich2009}. We analyze both full 
Lindbladian evolution and a unitary-quench protocol in which the system is first prepared in $\rho_{\mathrm{SS}}$ and then evolved with $\mathcal{L}_0(\rho)\equiv-i[H,\rho]$ (see SM~\cite{SM} for details).

For numerics, Eqs.~\eqref{eq:QGF} are well suited because they reduce the problem to a small number of time evolutions with modified 
initial conditions. 
In a vectorized formulation, one represents the density matrix as a state in Liouville space, $\dket{\rho}$, and the 
Liouvillian as a matrix-product operator acting on the doubled Hilbert space~\cite{Prosen.2008,Dzhioev.2011,Weimer2021,Moca.2022}. 
For a given $\lambda$, we prepare the twisted initial operator $\rho_\lambda(0)=e^{-i\lambda\Sigma_\ell}\rho_{\mathrm{SS}}$ (an MPO of low bond 
dimension), propagate it with the time-evolving block decimation (TEBD) algorithm to $\rho_\lambda(t)=e^{\mathcal{L}t}(\rho_\lambda(0))$, and evaluate $G_\ell(\lambda,t)=\mathrm{Tr}[e^
{i\lambda\Sigma_\ell}\rho_\lambda(t)]$ by tensor contraction~\cite{fishman2022itensor}. 
Cumulants can be extracted either by derivatives of $\ln G_\ell$ at 
$\lambda=0$ or, in practice, by a controlled small-$|\lambda|$ evaluation; using two complex phases of $\lambda$ (e.g., real and imaginary) 
cancels the leading truncation errors and stabilizes the extraction of $\kappa_2$~\cite{Valli.2025}.

\paragraph*{FV in the non-interacting limit.\textemdash}
\begin{figure*}[tbh!]
	\begin{center}
	 \includegraphics[width=0.62\columnwidth]{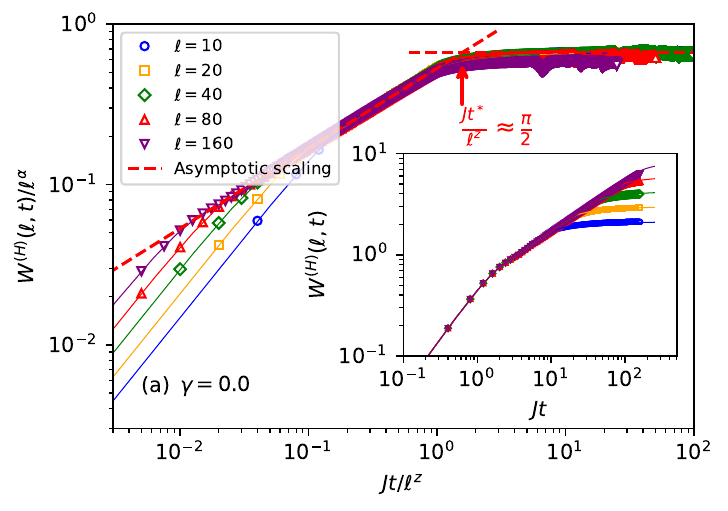}
	 \includegraphics[width=0.6\columnwidth]{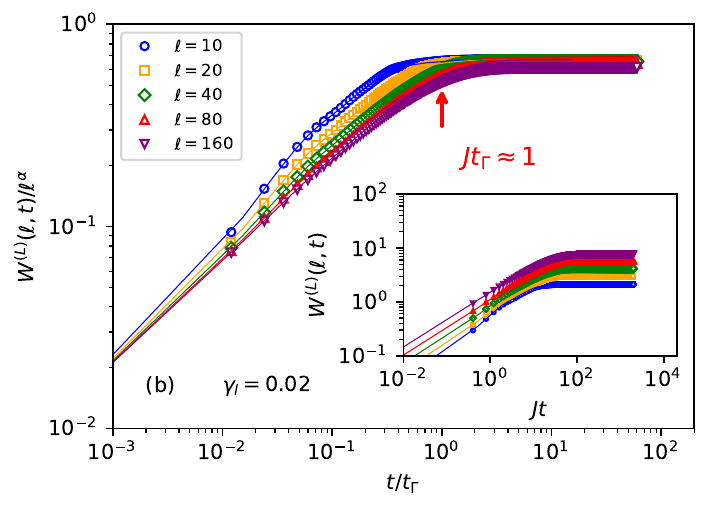}
	 \includegraphics[width=0.6\columnwidth]{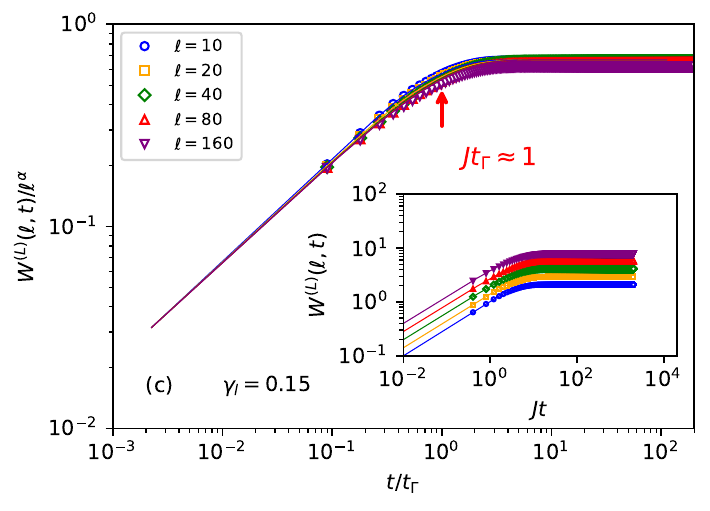}
	 \caption[FV scaling collapse, XX limit]{(a) FV scaling collapse for the unitary evolution in the XX limit ($\Delta=0$)
	 starting from the NESS with $\zeta=0.5$ ($1/3$-filling, $\bar n=1/3$). The symbols show the numerical TEBD data for
	 $W^{(H)}(\ell,t)$ while the solid lines show the analytic result from Eq.~\eqref{eq:kappa2_xx_finiteGamma} for the same
	 parameters. Dashed lines indicate the universal functions given in Eq.~\eqref{eq:fH_asymptotics}.
	  The crossover time $t^*$ is estimated from Eq.~\eqref{eq:tstar_unitary} and indicated by the vertical
	 arrow. (b,c) FV scaling collapse for the full Lindblad evolution for weak (b), ($\gamma_l=0.02J$) and 
	 strong (c) ($\gamma_l=0.15J$), dissipation respectively, and $\zeta=0.5$ as before. The roughness saturates already at $t\sim t_{\Gamma}$.  The crossover time $t_{\Gamma}$ is indicated by the vertical arrow. 
	 System sizes are $L=500$ and segment lengths
	 $\ell=10, 20,40,80, 160$. 
	 In the TEBD simulations, we used a maximum bond dimension $\chi_{\rm max}=64$ and a
	 time step $dt=0.01/J$. 
	 The inset shows the same unrescaled data, on a log-log scale, where the three regimes of microscopic transient, FV growth, and FV saturation are clearly visible.
	 }
	 \label{fig:FV_Delta_0}
	\end{center}
\end{figure*}
In the XX limit ($\Delta=0$), Jordan--Wigner fermionization~\cite{Lieb1961} maps $\Sigma_\ell$ to the segment particle number
 (up to a constant), such that $\Delta\Sigma_\ell(t)$ counts transferred fermions. 
For homogeneous gain/loss the model is quadratic and can be solved exactly, e.g., by third quantization~\cite{Prosen.2008,Prosen2010}. 
The second cumulant at filling $\bar n=\zeta/(1+\zeta)$ writes as
\begin{equation}
\kappa_2(\ell,t)=2\bar n(1-\bar n)\left[\ell-e^{-\Gamma t}\sum_{r=-(\ell-1)}^{\ell-1}(\ell-|r|)\,J_r^2(Jt)\right],
\label{eq:kappa2_xx_finiteGamma}
\end{equation}
with $J_r(x)$ Bessel functions and a ballistic kernel inherited from the unitary XX chain~\cite{Antal1999,Abramowitz1965}. Equation~\eqref{eq:kappa2_xx_finiteGamma} is our
  key results: the coherent light-cone structure survives, but the 
two-time correlations that encode memory of the initial segment configuration are exponentially damped by dissipation. 
In fact, for bulk observables in a translation-invariant geometry, one may view gain/loss as an effective dephasing/relaxation mechanism for 
dynamical fluctuations on top of $\rho_{\mathrm{SS}}$, which amounts to a replacement $|U_{ij}(t)|^2\to e^{-\Gamma t}|U_{ij}(t)|^2$ in connected density correlators,
$\langle n_i(t)n_j\rangle_{\mathrm{SS}}-\bar n^2=\bar n(1-\bar n)e^{-\Gamma t}|U_{ij}(t)|^2$ (see SM~\cite{SM} for a derivation).

This structure immediately clarifies the relevant time scales and universal functions. In the unitary-quench limit ($\Gamma\to0$) one obtains a one-parameter FV collapse with 
the ballistic scaling variable $x\equiv Jt/\ell$ and $z=1$, $W^{(H)}(\ell,t)=\ell^{1/2} f_H(x)$. This was observed experimentally in Ref. \cite{kwon.2026}.
After a short microscopic transient ($Jt\ll1$) where $\kappa_2\propto t^2$ and $W\propto t$, the pre-saturation window $1\ll Jt\ll \ell$ yields universal growth and a plateau,
\begin{equation}
f_H(x)\simeq
\begin{cases}
\left(\frac{4}{\pi}\bar n(1-\bar n)\,x\right)^{\!1/2}, & x\ll 1,\\
\big(2\bar n(1-\bar n)\big)^{\!1/2}, & x\gg 1,
\end{cases}
\label{eq:fH_asymptotics}
\end{equation}
as extracted from the limits of the Bessel functions. 
The crossover between growth and saturation is set by the coherent finite-size time $t^*\sim \ell/J$.
The unitary limit is recovered as $\Gamma\to0$ in Eq.~\eqref{eq:kappa2_xx_finiteGamma}, and the FV crossover time $t^*$ can be estimated as
\begin{equation}
 \frac{Jt^*}{\ell}\simeq \frac{\pi}{2}.
\label{eq:tstar_unitary}
\end{equation}
In the full Lindblad dynamics, dissipation introduces a second dimensionless parameter $y\equiv\Gamma t$ in addition to $x=Jt/\ell$, so the natural description is a two-parameter scaling form $W^{(\mathcal{L})}(\ell,t)=\ell^{1/2}f_{\mathcal{L}}(x,y)$, with the unitary FV function recovered as $f_{\mathcal{L}}(x,0)=f_H(x)$. Equation~\eqref{eq:kappa2_xx_finiteGamma} implies the asymptotics
\begin{equation}
f_{\mathcal{L}}(x,y)\simeq
\begin{cases}
\left(\frac{4}{\pi}\bar n(1-\bar n)\,x\right)^{\!1/2}, & x\ll 1\ \text{and}\ y\ll 1,\\
\big(2\bar n(1-\bar n)\big)^{\!1/2}, & x\gg 1\ \text{or}\ y\gg 1,
\end{cases}
\label{eq:fL_asymptotics}
\end{equation}
namely unitary-like growth for $y\ll1$ and saturation once either finite-size decorrelation ($x\gg1$) or dissipation-induced loss of memory ($y\gg1$) sets in. Accordingly, besides the coherent time $t^*\sim\ell/J$, there is a dissipative crossover time $t_\Gamma\sim\Gamma^{-1}$, and the unitary-like FV growth regime requires $t\ll\min\{t^*,t_\Gamma\}$.

Two limiting scenarios are particularly relevant: (\emph{i}) In the weak-dissipation regime $t_\Gamma\gg t^*$ (equivalently $\Gamma\ll J/\ell$), the system explores 
the full ballistic FV scaling before dissipation acts, and a one-parameter collapse in $x$ is observed over a broad range;  (\emph{ii}) In the opposite, 
dissipation-dominated regime $t_\Gamma\ll t^*$ (i.e., $\Gamma\gg J/\ell$), the roughness saturates already at $t\sim t_\Gamma$ while $x=Jt/\ell\ll1$, so a 
one-parameter collapse in $x$ necessarily breaks down beyond very short times; instead, the approach to the plateau is governed primarily by $y=\Gamma t$ at fixed small $x$, consistent with the finite-$\Gamma$ panels in Fig.~\ref{fig:FV_Delta_0}.

Figure~\ref{fig:FV_Delta_0} illustrates these two regimes in a single dataset by tuning $\Gamma$. Panel (a) shows the unitary-quench limit ($\Gamma=0$), where 
the FV collapse is controlled by the coherent scaling variable $x=Jt/\ell$ and the analytic Bessel-kernel result captures both the growth branch and the 
finite-size saturation. Panels (b) and (c) show the full Lindblad evolution for weak and strong dissipation, respectively: for smaller $\Gamma$ as in panel (b) the curves 
still follow the unitary-like collapse at early times $t\ll t_\Gamma$, while the eventual saturation is shifted to the dissipative crossover scale $t\sim t_\Gamma$; for larger $\Gamma$ as in panel (c) the dissipative relaxation sets in before the coherent finite-size time is reached, and the system is rapidly driven into 
the dissipation-dominated saturation regime even though $Jt/\ell$ remains small. 
In each panel the symbols show the numerical TEBD data, while the solid lines show the analytical result from Eq.~\eqref{eq:kappa2_xx_finiteGamma} for the same parameters. 

\paragraph*{Interacting case.\textemdash}

\begin{figure}[tbh!]
	\centering
	\begin{center}
	 \includegraphics[width=0.9\columnwidth]{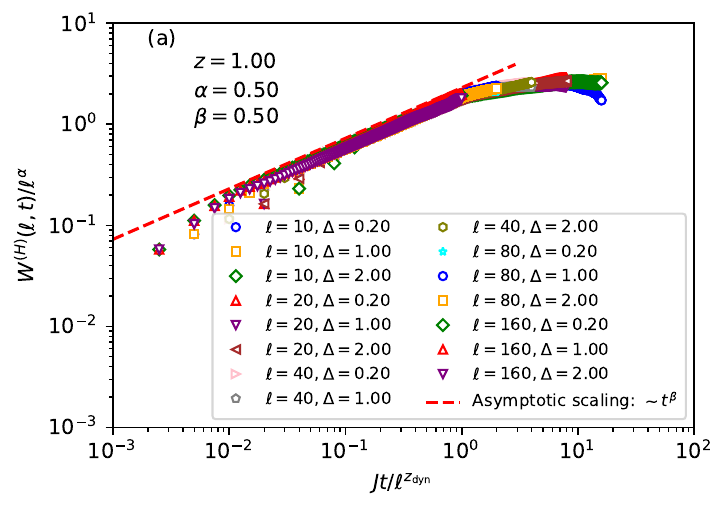}
	 \includegraphics[width=0.9\columnwidth]{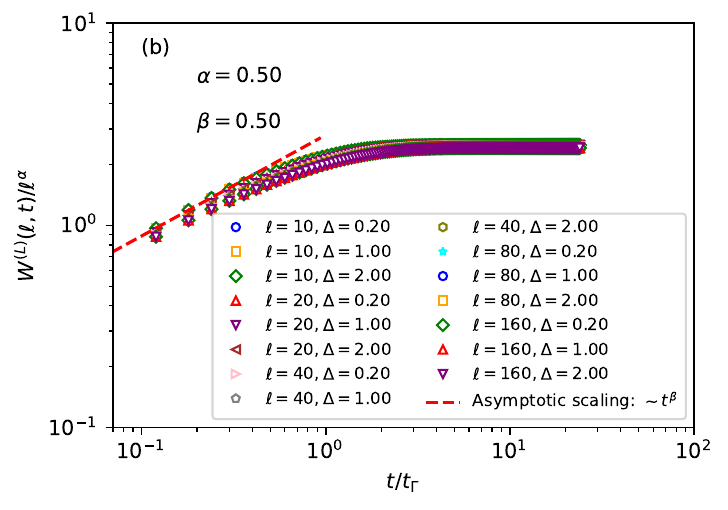}
	 \caption[Breakdown of FV collapse under dissipation]{(a) FV scaling collapse for the unitary evolution in the interacting XXZ
	chain with $\Delta=\{0.2, 1.0, 2.0\}$ and $\zeta=0.1$. The symbols show the numerical TEBD data while the
	dashed  lines are the asymptotic forms of the universal function corresponding to the growth $\sim t^{\beta}$.
	(b) same as in (a) but for the Lindbladian evolution with $\gamma_l=0.15 J$ and $\zeta=0.1$.
	System sizes are $L=500$ and segment lengths $\ell=10, 20,40,80, 160$. 
	In the TEBD simulations, we use maximum bond dimension $\chi_{\rm max}=32$ and a time step $dt=0.01/J$.}
	 \label{fig:FV_interacting}
	\end{center}
\end{figure}

In the interacting XXZ chain ($\Delta\neq0$), closed analytic expressions for the QGF and the resulting roughness are not available, so we 
extract FV scaling from numerical time evolution using the same protocol as in the noninteracting case: we prepare the chain in the diagonal 
product NESS $\rho_{\mathrm{SS}}(\zeta)$ and evolve it either (i) unitarily with $H$ or (ii) with the full Liouvillian. The roughness is 
obtained from the second cumulant of the transferred segment magnetization, $W(\ell,t)=\sqrt{\kappa_2(\ell,t)}$, computed via the vectorized 
TEBD evaluation of the QGF (details in \cite{SM}). Because $\rho_{\mathrm{SS}}$ is a product state, the subsystem-size 
dependence of the plateau remains consistent with $W_{\rm sat}(\ell)\propto \ell^{\alpha}$ with $\alpha\simeq1/2$; the dynamical information 
is therefore encoded primarily in the growth regime and in the crossover time scales.

Figure~\ref{fig:FV_interacting}(a) demonstrates that at finite magnetization density ($\zeta=0.1$) the unitary dynamics exhibits a robust 
one-parameter FV collapse across anisotropies $\Delta=\{0.2,1.0,2.0\}$ when plotted in the ballistic variables, i.e., with a single scaling 
variable $x\sim t/\ell$ and an effective dynamical exponent compatible with $z\simeq1$. Interactions mainly renormalize nonuniversal 
amplitudes and the microscopic transient, while the overall collapse remains intact, consistent with a quasiparticle (magnon-like) picture at finite polarization in which magnetization fluctuations propagate at a finite velocity~\cite{Schmitteckert.2020}. The dashed lines in Fig.
~\ref{fig:FV_interacting}(a) indicate that the growth branch is compatible with a power law $W\propto t^{\beta},(\beta=1/2)$ over the pre-saturation window.

At the infinite-temperature point $\zeta=1$, the unitary dynamical exponent depends on $\Delta$ (ballistic for $|\Delta|<1$, KPZ-type $z\simeq3/2$ at $\Delta=1$, and diffusive $z\simeq2$ for $\Delta>1$), consistent with known XXZ transport regimes~\cite{Znidaric.2011,Ljubotina.2017,Ljubotina_2017,Marko.2019,Bulchandani.2019,DeNardis2019,Bulchandani_2021,Gopalakrishnan2023}.

Figure~\ref{fig:FV_interacting}(b) shows that adding homogeneous gain and loss qualitatively changes the scaling structure: the roughness saturates already at times $t\sim t_\Gamma$ even when $Jt/\ell\ll1$, so a one-parameter collapse as a function of the coherent FV variable $t/\ell^{z}$ generically fails once the dissipative relaxation sets in. Consistently, the data are instead governed primarily by the microscopic dissipative scale $t_\Gamma\sim \Gamma^{-1}$, and the curves collapse best when time is measured in units of $y\equiv\Gamma t$ (with interactions and $\Delta$ affecting mainly the early-time coherent transient). 
This behavior is naturally captured by a two-parameter scaling form $W^{(\mathcal{L})}(\ell,t)=\ell^{1/2} f_{\mathcal{L}}(Jt/\ell,\Gamma t)$, with a dissipation-dominated regime in which $W^{(\mathcal{L})}(\ell,t)\simeq \ell^{1/2} g(\Gamma t)$ (numerical details are presented in \cite{SM}).

\paragraph*{Integrability breaking.\textemdash}
The robustness of FV scaling away from the integrable XXZ limit is an important question, since generic many-body systems are non-integrable. We therefore add a next-nearest-neighbor term
$H_{\mathrm{nnn}}=(J_2/4)\sum_l(\sigma_l^x\sigma_{l+2}^x+\sigma_l^y\sigma_{l+2}^y+\Delta\,\sigma_l^z\sigma_{l+2}^z)$,
which breaks integrability for $\Delta\neq0$ while preserving the $\mathrm{U}(1)$ symmetry. Crucially for our purposes, this perturbation leaves the diagonal product NESS $\rho_{\mathrm{SS}}(\zeta)$ unchanged; thus the plateau scaling $W_{\rm sat}(\ell)\propto \ell^{\alpha}$ with $\alpha\simeq 1/2$ is fixed by the initial-state variance and provides a stable reference for comparing coherent growth and crossover scales.

For the unitary-quench protocol at finite magnetization density ($\zeta\neq1$), the FV collapse remains compatible with a ballistic scenario 
even after integrability is broken: the rescaled curves are still well organized by a single coherent scaling variable $t/\ell$ and a 
crossover time $t^*\propto\ell$. The main effect of $J_2$ is quantitative---it renormalizes the effective propagation velocity (and more 
generally the distribution of quasiparticle velocities), which shifts the location of the crossover without destroying the collapse in the 
explored time window. This picture is particularly clear in the exactly solvable limit $\Delta=0$ with $J_2\neq0$, 
where the model remains quadratic and one can explicitly verify that the FV exponents stay ballistic ($z=1$) while the scaling variable is 
controlled by an effective velocity set by the modified dispersion (details in \cite{SM}). By contrast, at the infinite-temperature point $\zeta=1$, integrability breaking is known to promote diffusion over a broad range of $\Delta$~\cite{Moca.2026}, consistent with a change in 
the coherent dynamical exponent inferred from FV collapse.

For the full Lindbladian evolution, non-integrability does not restore a coherent one-parameter FV collapse in $Jt/\ell$: the roughness 
typically saturates already at $t\sim t_\Gamma\sim\Gamma^{-1}$, i.e., before the coherent finite-size time $t^*$ is reached, so the dynamics 
is governed primarily by the dissipative scaling variable $\Gamma t$. 
In this sense, bulk gain/loss acts as a channel that dominates the loss of memory required by the QGF-based FV construction, largely independent of whether the underlying Hamiltonian dynamics is integrable.

\paragraph*{Conclusions.\textemdash}
We have shown that Family--Vicsek scaling provides a compact way to organize 
magnetization-fluctuation dynamics in a quenched dissipative XXZ chain. 
The key message is that FV scaling survives in open quantum systems: coherent spreading can 
still generate a FV growth-and-saturation pattern, while local gain/loss introduces an additional microscopic relaxation that can 
dominate the crossover.

In the noninteracting XX limit, this competition can be resolved analytically and makes clear why dissipation changes the scaling structure: 
ballistic propagation remains visible, but the  memory required for transferred-magnetization fluctuations is gradually erased, so 
saturation is ultimately set by the dissipative time scale. 
In the interacting chain, numerical TEBD indicates that unitary FV collapse is 
robust at finite magnetization density across a broad range of anisotropies, consistent with finite-velocity transport of fluctuations. 
By contrast, under Lindbladian evolution the dominant organization of the dynamics is generically governed by the dissipation rate, and a 
coherent one-parameter collapse in the transport variable is not sustained.

Finally, we found that these qualitative conclusions are stable under integrability breaking by next-nearest-neighbor couplings: at $\zeta\neq1$ the unitary ballistic scenario persists with renormalized velocity scales, whereas the dissipation-dominated collapse remains 
largely insensitive to integrability. 
The supplementary material \cite{SM} provides technical details and additional 
 supporting data.

\begin{acknowledgments}
This work received financial support from CNCS/CCCDI-UEFISCDI, under projects number PN-IV-P1-PCE-2023-0159 and PN-IV-P1-PCE-2023-0987 and was also 
supported by the National Research, Development and Innovation Office - NKFIH  Project No. K142179.
We acknowledge the Digital Government Development and Project Management Ltd.~for awarding us access to the Komondor HPC facility based in Hungary.

\end{acknowledgments}

\bibliography{references}

\clearpage
\onecolumngrid

\section*{Supplemental Material}
This Supplementary Material provides further derivations, numerical details, and supporting results for the Family--Vicsek (FV) scaling analysis reported there.
We summarize the main physical picture: FV scaling provides a compact, universal description of how fluctuations grow, spread, and saturate in time, and we extend it to magnetization fluctuations after quenches in a driven--dissipative spin-$1/2$ XXZ chain with bulk gain and loss.
Starting from the nonequilibrium steady state, we compute the roughness from the second cumulant of the transferred magnetization of a finite subsystem.
In the noninteracting XX limit under unitary evolution, we identify three temporal regimes analytically: a short-time microscopic transient, followed by universal FV growth and FV saturation, corresponding to a ballistic universality class.
For the full (non-unitary) Lindblad dynamics, saturation is governed by the dissipative time scale.
In the presence of interactions, we find ballistic transport for magnetic initial states for all values of the interaction parameter due to ballistic magnon propagation, while the interacting Lindbladian dynamics is dominated by a dissipation-controlled collapse.

\begingroup
\printsupplementaltableofcontents
\endgroup

\clearpage

\section{Introduction}
Our focus is on the time-dependent growth and saturation of magnetization fluctuations after quenches in a driven--dissipative spin-$1/2$ XXZ chain with bulk gain and loss, described by a Markovian Lindblad master equation~\cite{Gorini.1976,Lindblad.1976}.

We characterize fluctuations in a segment of length $\ell$ through the roughness $W(\ell,t)=\sqrt{\kappa_2(\ell,t)}$, extracted from the second cumulant of the transferred segment magnetization.
The scaling analysis is organized using the Family--Vicsek (FV) scaling ansatz~\cite{family1985scaling}:
\begin{equation}
W(\ell,t) \sim \ell^{\alpha}\, f\!\left(\frac{t}{\ell^{z}}\right),
\label{sm:eq:FV_scaling}
\end{equation}
with $f(x\ll 1)\sim x^{\beta}$ and $f(x\gg 1)\sim \mathrm{const}$, and exponents related by $z=\alpha/\beta$.
Here $z$ is the dynamical exponent, $\alpha$ describes the subsystem-size dependence of the saturated roughness, and $\beta$ characterizes the early-time growth.

Two technical aspects are especially relevant for the driven--dissipative setting: (i) the initial state is the exactly known nonequilibrium steady state (a diagonal product state parameterized by $\zeta$-the fugacity of the state), and (ii) the roughness relevant for FV scaling is naturally encoded in a two-time, counting-field generating function rather than in a single-time Hermitian observable.
The purpose of this Supplementary Material is to make these constructions explicit and to provide analytic benchmarks (in the XX limit) alongside the numerical protocol.

\section{Driven--dissipative XXZ model and nonequilibrium steady state}
\label{sm:sec:model_ness}
We study a spin-$1/2$ XXZ chain of length $L$ subject to homogeneous local gain and loss.
The system is treated as a Markovian open quantum many-body system whose density matrix $\rho(t)$ evolves according to a Lindblad master equation~\cite{Gorini.1976,Lindblad.1976},
\begin{align}
\partial_t\rho(t) &= \mathcal L[\rho(t)]\nonumber\\
&= -i[H,\rho(t)] + \mathcal{D}[\rho(t)],
\end{align}
with $\mathcal L$ the Liouvillian superoperator governing the dynamics.
The coherent part is generated by the XXZ Hamiltonian
\begin{equation}
H = \frac{J}{4} \sum_{l=-L/2}^{L/2-1}
\left(
\sigma_l^x \sigma_{l+1}^x
+ \sigma_l^y \sigma_{l+1}^y
+ \Delta \, \sigma_l^z \sigma_{l+1}^z
\right),
\end{equation}
where $\sigma_l^{\alpha}$ ($\alpha=x,y,z$) are Pauli matrices at site $l$, $J$ sets the energy scale, and $\Delta$ is the anisotropy.

The dissipative part consists of on-site incoherent spin flips, described by the jump operators
\begin{equation}
L_l^{-} = \sqrt{\gamma_l}\,\sigma_l^{-},\qquad L_l^{+} = \sqrt{\gamma_p}\,\sigma_l^{+},
\end{equation}
with $\sigma_l^{\pm}=(\sigma_l^x\pm i\sigma_l^y)/2$.
Here $\gamma_l$ and $\gamma_p$ denote the loss and pumping rates, respectively.
The corresponding dissipator reads
\begin{gather}
\mathcal{D}[\rho]=\sum_{l=-L/2}^{L/2-1}
\Big[
\gamma_l\Big(\sigma_l^-\rho\sigma_l^+-\tfrac{1}{2}\{\sigma_l^+\sigma_l^-,\rho\}\Big)\\\nonumber
+\gamma_p\Big(\sigma_l^+\rho\sigma_l^--\tfrac{1}{2}\{\sigma_l^-\sigma_l^+,\rho\}\Big)
\Big].
\label{sm:eq:Dissipation}
\end{gather}
For $\gamma_p,\gamma_l>0$, the interplay of coherent XXZ dynamics and local gain/loss drives the chain to a unique nonequilibrium steady state (NESS)
defined by
\begin{equation}
\partial_t\rho_{\mathrm{SS}}=0.
\end{equation}
It is convenient to parameterize the steady state by the ratio
\begin{equation}
\zeta\equiv \frac{\gamma_p}{\gamma_l},
\end{equation}
which plays the role of a spin fugacity.
Remarkably, the NESS can be written in closed form and is independent of $J$ and $\Delta$.
In the $\sigma^z$ basis it is diagonal and coincides with an infinite-temperature state at fixed fugacity $\zeta$.
One representation is a weighted sum over magnetization sectors,
\begin{equation}
\rho_{\mathrm{SS}}=\frac{1}{{(1+\zeta)}^L}\sum_{M=0}^{L} \zeta^M\,\mathbb{1}_M,
\end{equation}
where $\mathbb{1}_M$ denotes the identity operator in the subspace with total magnetization $M$.
Equivalently, $\rho_{\mathrm{SS}}$ factorizes into a product of identical one-site density matrices,
\begin{equation}
\rho_{\mathrm{SS}}=\bigotimes_{l=-L/2}^{L/2-1}\rho_l,\qquad
\rho_l=\frac{1}{1+\zeta}
\begin{pmatrix}
\zeta & 0\\
0 & 1
\end{pmatrix}.
\label{sm:eq:rho_NESS}
\end{equation}
This product structure implies the absence of connected equal-time correlations in the NESS.

The local magnetization is uniform and given by
\begin{equation}
\langle \sigma_l^z \rangle_{\mathrm{SS}} = \frac{\zeta-1}{\zeta+1},
\label{sm:eq:Sz_NESS}
\end{equation}
while transverse coherences vanish, $\langle \sigma_l^{\pm} \rangle = 0$. Coherent exchange interactions do not affect 
the structure of the steady state, but only govern the transient dynamics and the relaxation towards stationarity.

Importantly, this conclusion is unchanged if one augments the coherent dynamics by additional interaction terms that conserve the total magnetization $S^z=\tfrac{1}{2}\sum_l\sigma_l^z$.
As an example, one may add a next-nearest-neighbor coupling,
\begin{equation}
H_{\rm nnn}=\frac{J_2}{4}\sum_{l=-L/2}^{L/2-2}
\left(\sigma_l^x\sigma_{l+2}^x+\sigma_l^y\sigma_{l+2}^y+\Delta\,\sigma_l^z\sigma_{l+2}^z\right),
\label{sm:eq:H_nnn}
\end{equation}
and consider $H\to H+H_{\rm nnn}$.
Since $[H_{\rm nnn},S^z]=0$, the steady state in Eq.~\eqref{sm:eq:rho_NESS} can be viewed as a function of the good quantum number $S^z$ and therefore commutes with the full Hamiltonian.
Consequently, the NESS remains entirely controlled by the dissipator $\mathcal{D}$ and retains the same product form; the extra coherent term only modifies the transient (post-quench) dynamics.

We have verified numerically (using a vectorized time-evolving block decimation (TEBD)~\cite{Vidal.2003,Vidal.2004,Daley.2004} implementation using the \texttt{ITensor.jl} library~\cite{fishman2022itensor}) that the steady-state magnetization matches Eq.~\eqref{sm:eq:Sz_NESS} across the parameter range considered.
This provides a direct check that the long-time state is fixed solely by the local gain/loss balance, while $J$ and $\Delta$ control only the approach to stationarity.
For the particular case $\zeta=1$, the NESS density matrix 
corresponds to the infinite temperature state, $\rho_{\mathrm{SS}}\propto \mathbb{1}$, and the average magnetization vanishes, $\average{\sigma^{z}_l}_{\mathrm{SS}}=0$.  
While static correlations are fully suppressed, nontrivial physics emerges in the time-dependent correlation functions 
evaluated on top of this steady state, which probe how coherent spin excitations propagate and decay in the presence of 
local dissipation.

\section{Vectorized Quantum Generating Function approach}
\label{sm:sec:QGF}

To access the full time-dependent fluctuations that enter the FV scaling analysis, we use a generating-function formulation that efficiently yields cumulants of extensive observables without sampling the full probability distribution~\cite{Moca.2026}.
We focus on the magnetization in a segment of length $\ell$,
\begin{equation}
\Sigma_\ell \equiv \frac{1}{2}\sum_{j\in \mathrm{seg}(\ell)} \sigma_j^z,
\end{equation}

and on the corresponding transferred variable
\begin{equation}
\Gamma(t) \equiv \Sigma_\ell(t)-\Sigma_\ell(0),
\end{equation}

whose second cumulant sets the roughness via
\begin{equation}
W(\ell,t)=\sqrt{\kappa_2(\ell,t)}.\label{sm:eq:roughness_W}    
\end{equation}
The central object is the operator
\begin{equation}
R_\ell(\lambda)\equiv e^{i\lambda \Sigma_\ell},
\end{equation}
defined in terms of a (generally complex)  field $\lambda$.

Importantly, the quantity we ultimately need is intrinsically a \emph{two-time} object: it generates cumulants of the 
transferred magnetization $\Gamma(t)=\Sigma_\ell(t)-\Sigma_\ell(0)$ and therefore necessarily combines operators 
evaluated at times $0$ and $t$ (schematically, $G_\ell(\lambda,t)\sim\langle e^{i\lambda\Sigma_\ell(t)}e^{-i\lambda\Sigma_\ell(0)}\rangle$).
As in quantum work statistics~\cite{Talkner2007,Silva.2008} and in full counting statistics~\cite{Levitov1993,Klich2009}, such characteristic functions are not 
associated with a unique single-time Hermitian observable; their operational meaning depends on a specified measurement protocol or, equivalently, on introducing a counting-field $\lambda$ of the dynamics.
While this point is subtle, we find that this two-time construction is the natural---and, in practice, the only 
viable---way to capture FV roughness scaling in quantum spin systems.

In the Heisenberg picture of Lindblad dynamics, operators evolve as $O(t)=e^{\mathcal{L}^{\dagger} t}(O)$, where $\mathcal{L}$ is the Liouvillian superoperator appearing in the master equation.
In the following we compute two versions of the QGF, corresponding to (i) full Lindbladian evolution and  (ii) unitary 
evolution obtained by switching off dissipation after preparing $\rho_{\mathrm{SS}}$.
Concretely, we define
\begin{align}
G_\ell^{(\mathcal{L})}(\lambda,t)
&\equiv \mathrm{Tr}\!\left[ R_\ell(\lambda)\, e^{\mathcal{L} t}\!\left(R_\ell^\dagger(\lambda)\,\rho_{\mathrm{SS}}\right)\right],
\label{sm:eq:QGF_Lindblad}
\\
G_\ell^{(H)}(\lambda,t)
&\equiv \mathrm{Tr}\!\left[ R_\ell(\lambda)\, e^{\mathcal{L}_0 t}\!\left(R_\ell^\dagger(\lambda)\,\rho_{\mathrm{SS}}\right)\right]\nonumber \\
&= \mathrm{Tr}\!\left[ R_\ell(\lambda)\, e^{-iHt} R_\ell^\dagger(\lambda)\,\rho_{\mathrm{SS}}\, e^{iHt}\right].
\label{sm:eq:QGF_unitary}
\end{align}
Here $\mathcal{L}_0(\rho)\equiv -i[H,\rho]$ generates purely unitary dynamics. 
Formally, $G_\ell^{(X)}(\lambda,t)$ is the characteristic function associated with $\Gamma(t)$ (for a given dynamics $X\in\{\mathcal{L},H\}$), and the corresponding distribution can be obtained by Fourier transform,
\begin{equation}
P_\ell^{(X)}(\Gamma,t)=\frac{1}{2\pi}\int d\lambda\, e^{-i\lambda\Gamma}\,G_\ell^{(X)}(\lambda,t).
\end{equation}
In practice we are interested in cumulants,
\begin{equation}
\kappa_n^{(X)}(\ell,t)=(-i)^n\,\partial_\lambda^n\left.\ln G_\ell^{(X)}(\lambda,t)\right|_{\lambda=0},
\label{sm:eq:cumulants_def}
\end{equation}
from which $W^{(X)}(\ell,t)=\sqrt{\kappa_2^{(X)}(\ell,t)}$ follows for $n=2$.
We analyze FV scaling separately for each dynamics,
\begin{align}
W^{(X)}(\ell,t) &\sim \ell^{\alpha_{X}}\, f_{X}\!\left(\frac{t}{\ell^{z^{(X)}}}\right),
\label{sm:eq:FV_unitary}
\end{align}
For sufficiently small $|\lambda|$, one may expand
\begin{equation}
G_\ell^{(X)}(\lambda,t)=1-\frac{\lambda^2}{2}\,\mu_2^{(X)}(\ell,t)+\mathcal{O}(\lambda^4),
\end{equation}
where $\mu_2^{(X)}(\ell,t)=\langle \Gamma(t)^2\rangle_X$ if odd moments vanish by symmetry (as in the $\zeta=1$ steady state, where the magnetization distribution is centered).
This yields the second moment and cumulant in the small-$\lambda$ limit,
\begin{equation}
\kappa_2^{(X)}(\ell,t)=\mu_2^{(X)}(\ell,t)\simeq \frac{2}{\lambda^2}\big(1-\mathrm{Re}\,G_\ell^{(X)}(\lambda,t)\big)+\mathcal{O}(\lambda^2).
\label{sm:eq:kappa2_small_lambda}
\end{equation}
To suppress the leading $\mathcal{O}(\lambda^2)$ truncation error, it is advantageous to evaluate $G_\ell$ at two phases of the complex field and combine them~\cite{Valli.2025}.
For example, choosing $\lambda=r$ and $\lambda=i r$ (with $r\in\mathbb{R}$) cancels the $\mathcal{O}(r^2)$ contribution,
\begin{equation}
\mu_2^{(X)}(\ell,t)\simeq \frac{G_\ell^{(X)}(i r,t)-G_\ell^{(X)}(r,t)}{r^2}+\mathcal{O}(r^4),
\label{sm:eq:mu2_phase_combo}
\end{equation}
and analogous phase combinations allow one to access higher moments and cumulants if needed.
For $\zeta\neq 1$, one can either work with the centered operator $\delta\Sigma_\ell\equiv \Sigma_\ell-\langle\Sigma_\ell\rangle_{\mathrm{SS}}$ (replacing $\Sigma_\ell$ in $R_\ell$) or evaluate $\kappa_1$ and $\kappa_2$ separately from Eq.~\eqref{sm:eq:cumulants_def}.

Equations~\eqref{sm:eq:QGF_Lindblad}--\eqref{sm:eq:QGF_unitary} are particularly well suited to our numerics because they reduce the 
problem to a small number of time evolutions with modified initial conditions.
In a vectorized TEBD implementation, the density matrix is represented as a matrix-product state in Liouville space, $\dket{\rho}$, 
and the Liouvillian is represented as a matrix-product operator acting on the doubled Hilbert space~\cite{Prosen.2008,Dzhioev.2011,Weimer2021,Moca.2022}.
For each chosen counting field $\lambda$, we prepare the twisted initial state $\rho_\lambda(0)=R_\ell^\dagger(\lambda)\,\rho_{\mathrm{SS}}$ (an MPO with minimal bond dimension), evolve it according to $\rho_\lambda(t)=e^{\mathcal{L} t}(\rho_\lambda(0))$ using TEBD 
in Liouville space, and finally evaluate the overlap
\begin{equation}
G_\ell^{(\mathcal{L})}(\lambda,t)=\mathrm{Tr}\big[ R_\ell(\lambda)\,\rho_\lambda(t)\big]
\end{equation}
by contracting the corresponding MPO/MPDO tensors.
Because $R_\ell(\lambda)$ is close to the identity for small $|\lambda|$ and is supported only on the segment, the operator 
entanglement generated in $\rho_\lambda(t)$ remains comparable to that of the physical evolution from $\rho_{\mathrm{SS}}$, so that 
cumulants can be extracted at a cost similar to standard time propagation.

\section[Family-Vicsek scaling, Delta=0]{Family-Vicsek scaling in the non-interacting limit \texorpdfstring{$\Delta=0$}{Delta=0}}	
\label{sm:sec:FV_XX}
\subsection{Unitary evolution}

The unitary dynamics discussed above corresponds to the following quench protocol: the system is first prepared in the nonequilibrium steady state $\rho_{\mathrm{SS}}$ of the driven--dissipative problem, and subsequently the dissipator is switched off and the time evolution proceeds under the closed XXZ Hamiltonian $H$.
Since $\rho_{\mathrm{SS}}\propto e^{\mu S^z}$ and $[H,S^z]=0$, the state remains stationary under the unitary evolution; nevertheless, nontrivial dynamical fluctuations are encoded in the two-time object $G_\ell^{(H)}(\lambda,t)$ and its cumulants.

\label{sm:sec:unitary_mu2_XX}
In the noninteracting limit $\Delta=0$, the XXZ Hamiltonian reduces to the XX chain, which can be mapped exactly to spinless fermions via the Jordan--Wigner transformation~\cite{Lieb1961}.
We introduce fermionic operators $c_j$ satisfying $\{c_i,c_j^\dagger\}=\delta_{ij}$ and use
\begin{equation}
\sigma_j^+ = c_j^\dagger\exp\!\left(i\pi\sum_{m<j} c_m^\dagger c_m\right),\qquad
\sigma_j^z = 2 c_j^\dagger c_j-1.
\label{sm:eq:JW}
\end{equation}
In fermionic language, the Hamiltonian becomes
\begin{equation}
H=\frac{J}{2}\sum_{j}\left(c_j^\dagger c_{j+1}+\mathrm{h.c.}\right)
+ J\Delta\sum_{j}\left(n_j-\tfrac{1}{2}\right)\left(n_{j+1}-\tfrac{1}{2}\right),
\label{sm:eq:H_fermions}
\end{equation}
with $n_j\equiv c_j^\dagger c_j$.
Setting $\Delta=0$ yields a quadratic hopping problem with single-particle dispersion $\epsilon(k)=J\cos k$.
The segment magnetization is related to the fermion number in the segment,
\begin{equation}
\Sigma_\ell = \frac{1}{2}\sum_{j\in\mathrm{seg}(\ell)}\sigma_j^z
=\sum_{j\in\mathrm{seg}(\ell)}\left(n_j-\tfrac{1}{2}\right)
\equiv N_\ell-\frac{\ell}{2},
\end{equation}
so that the transferred variable simplifies to
\begin{equation}
\Gamma(t)=\Sigma_\ell(t)-\Sigma_\ell(0)=N_\ell(t)-N_\ell(0).
\label{sm:eq:Gamma_is_number_transfer}
\end{equation}
The steady state $\rho_{\mathrm{SS}}$ corresponds to an infinite-temperature grand-canonical ensemble with uniform occupation,
\begin{equation}
\bar n\equiv \langle n_j\rangle_{\mathrm{SS}}=\frac{\zeta}{1+\zeta},
\label{sm:eq:nbar}
\end{equation}
independent of $j$. 
The second moment (which equals the second cumulant for the stationary state considered here) reads
\begin{align}
\mu_2^{(H)}(\ell,t)=\kappa_2^{(H)}(\ell,t)
&=\Big\langle\big[N_\ell(t)-N_\ell(0)\big]^2\Big\rangle_{\mathrm{SS}}\nonumber\\
&=2\Big(\langle N_\ell^2\rangle_{\mathrm{SS}}-\langle N_\ell(t)N_\ell\rangle_{\mathrm{SS}}\Big),
\label{sm:eq:mu2_unitary_basic}
\end{align}
where we used $\langle N_\ell(t)^2\rangle_{\mathrm{SS}}=\langle N_\ell^2\rangle_{\mathrm{SS}}$, which follows from the stationarity of $\rho_{\mathrm{SS}}$ under the unitary evolution (see Appendix~\ref{sm:app:unitary_mu2} for a proof). In the XX limit, the fermionic operators evolve linearly,
\begin{equation}
c_j(t)=\sum_m U_{jm}(t)\,c_m,
\label{sm:eq:cj_evolution}
\end{equation}
and in the thermodynamic limit (or for segments far from the boundaries) one may use the translation-invariant kernel~\cite{Antal1999},
\begin{equation}
U_{jm}(t)=i^{\,m-j}\,J_{m-j}(J t),
\label{sm:eq:U_bessel}
\end{equation}
with $J_r$ the Bessel function of the first kind.
Using Wick's theorem for the Gaussian state and evaluating Eq.~\eqref{sm:eq:mu2_unitary_basic} yields
\begin{align}
\mu_2^{(H)}(\ell,t)
&=2\bar n(1-\bar n)\left[\ell-\sum_{r=-(\ell-1)}^{\ell-1}(\ell-|r|)\,J_r^2(Jt)\right].
\label{sm:eq:mu2_unitary_bessel}
\end{align}
This analytic result provides a useful benchmark for our tensor-network calculations and makes explicit the ballistic, light-cone-like spreading of fluctuations in the unitary XX chain.
Using Eq.~\eqref{sm:eq:mu2_unitary_bessel}, we can extract the FV scaling coefficients and the crossover time separating the growth and saturation regimes.
\begin{figure}[tbh!]
	\begin{center}
	 \includegraphics[width=0.4\columnwidth]{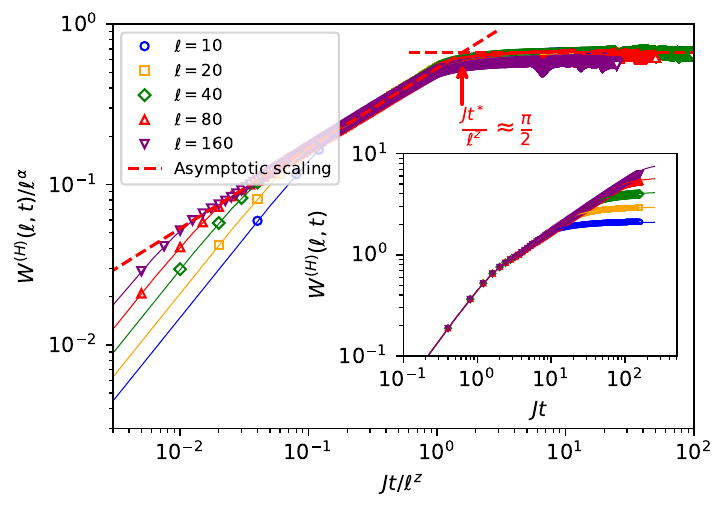}
	 \caption[FV scaling collapse, XX limit]{FV scaling collapse for the unitary evolution in the XX limit ($\Delta=0$) 
	 starting from the NESS with $\zeta=0.5$ ($1/3$-filling, $\bar n=1/3$). The symbols show the numerical TEBD data for 
	 $W^{(H)}(\ell,t)$ while the solid lines show the analytic result from Eq.~\eqref{sm:eq:mu2_unitary_bessel} for the same 
	 parameters. The crossover time $t^*$ is estimated from Eq.~\eqref{sm:eq:tstar_unitary} and indicated by the vertical  
	 arrow. The universal function  is shown as dashed lines, with the growth and saturation regimes corresponding to 
	 Eqs.~\eqref{sm:eq:W_unitary_growth}   and \eqref{sm:eq:W_unitary_sat}. System sizes are $L=1000$ and segment lengths 
	 $\ell=10, 20,40,80, 160$ and in the TEBD simulations we used a maximum bond dimension $\chi_{\rm max}=64$ and a 
	 time step $dt=0.01/J$. The agreement between the numerical data and the analytic result is excellent, confirming 
	 the validity of the FV scaling collapse and the extracted exponents. The inset shows the same unrescaled data, on a 
	 log-log scale, where the three regimes of microscopic transient, FV growth, and FV saturation are clearly visible.
	 }
	 \label{sm:fig:FV_unitary_Delta_0}
	\end{center}
\end{figure}

At very short times, $Jt\ll 1$, the roughness shows a microscopic transient that precedes the FV growth regime.
This follows directly from the small-argument expansion of the Bessel functions in Eq.~\eqref{sm:eq:mu2_unitary_bessel},
$J_0(x)\simeq 1-x^2/4$, $J_{\pm 1}(x)\simeq \pm x/2$, with $x\equiv Jt$.
Keeping the leading nontrivial contribution gives
\begin{equation}
\mu_2^{(H)}(\ell,t)\simeq \bar n(1-\bar n)(Jt)^2,\qquad Jt\ll 1,
\end{equation}
i.e., the second cumulant grows quadratically in time ($\kappa_2\propto t^2$), while the roughness grows linearly, $W^{(H)}(\ell,t)\propto t$.
This short-time behavior is consistent with the numerical data, which indicate that the FV scaling regime discussed below sets in only once $t\gtrsim 1/J$.

In the pre-saturation regime $1\ll Jt\ll \ell$, the Bessel weight $J_r^2(Jt)$ is concentrated at $|r|\lesssim Jt$ and one may approximate~\cite{Abramowitz1965}
\begin{equation}
\ell-\sum_{r=-(\ell-1)}^{\ell-1}(\ell-|r|)\,J_r^2(Jt)\simeq \sum_{r=-\infty}^{\infty}|r|\,J_r^2(Jt)\simeq \frac{2}{\pi}Jt,
\end{equation}
where the last asymptotics is the large-$t$ form of the mean displacement of the continuous-time quantum walk with probability distribution $p_r(t)=J_r^2(Jt)$. This yields
\begin{equation}
W^{(H)}(\ell,t)\simeq \left(\frac{4}{\pi}\bar n(1-\bar n)\,Jt\right)^{\!1/2},\qquad 1\ll Jt\ll \ell.
\label{sm:eq:W_unitary_growth}
\end{equation}
corresponding to FV growth with $\beta_H=1/2$. 
In the opposite limit $t\gg \ell/J$, the two-time correlator $\langle N_\ell(t)N_\ell\rangle_{\mathrm{SS}}$ decorrelates and Eq.~\eqref{sm:eq:mu2_unitary_basic} gives saturation to twice the static variance, $\mu_2^{(H)}(\ell,t\to\infty)\to 2\,\mathrm{Var}_{\mathrm{SS}}(N_\ell)=2\bar n(1-\bar n)\ell$, i.e.,
\begin{equation}
W^{(H)}_{\rm sat}(\ell)\simeq \big(2\bar n(1-\bar n)\big)^{\!1/2}\,\ell^{1/2},\qquad t\gg \ell/J.\,
\label{sm:eq:W_unitary_sat}
\end{equation}
corresponding to FV saturation with $\alpha_H=1/2$. Implictly, the dynamical exponent is $z^{(H)}=\alpha_H/\beta_H=1$, indicating a  ballistic universality class.
The crossover time $t^*$ separating the two asymptotic scaling forms can be estimated by matching $W^{(H)}(\ell,t^*)\simeq W^{(H)}_{\rm sat}(\ell)$, which gives
\begin{equation}
 \frac{Jt^*}{\ell}\simeq \frac{\pi}{2}.
\label{sm:eq:tstar_unitary}
\end{equation}
These expressions provide explicit estimates for the small-argument and large-argument amplitudes of the unitary FV scaling 
function when the scaling variable is taken as $x = Jt/\ell$, 
\begin{equation}
f_H(x) \simeq 
\begin{cases}
\left(\frac{4}{\pi}\bar n(1-\bar n)\,x\right)^{\!1/2}, & x\ll 1,\\
\big(2\bar n(1-\bar n)\big)^{\!1/2}, & x\gg 1.  
\label{sm:eq:fH_asymptotics}
\end{cases}
\end{equation}
Notice that the crossover time $t^*$ is independent of the filling 
$\bar n$ and is set solely by the coherent dynamics. 
Furthermore, the scaling function depends on $\bar n$ up 
to an overall multiplicative factor, which is a consequence of the Gaussian nature of the steady state and the linearity of 
the evolution.

Figure~\ref{sm:fig:FV_unitary_Delta_0} demonstrates that, for unitary dynamics in the XX limit, the roughness obeys 
a clean FV scaling collapse when rescaled with the ballistic variable $x=Jt/\ell$. 
The data for different segment lengths $\ell$ collapse onto a single scaling function with the growth and saturation 
branches set by Eqs.~\eqref{sm:eq:W_unitary_growth} and \eqref{sm:eq:W_unitary_sat}, corresponding to $\beta_H=1/2$ and $\alpha_H=1/2$ (hence $z^{(H)}=1$).
The vertical marker at $t^*$ highlights the finite-size crossover from the pre-saturation growth window ($1\ll Jt\ll \ell$) to saturation ($t\gg \ell/J$), while the inset makes visible the preceding microscopic transient.
The excellent agreement between the TEBD results (symbols) and the analytic free-fermion expression in Eq.~\eqref{sm:eq:mu2_unitary_bessel} (solid lines) provides a validation of both the numerical approach and the FV-exponent extraction. 

The inset, which shows the unrescaled roughness, makes the separation of time scales particularly 
transparent.
At the shortest times ($Jt\ll 1$) one observes a nonuniversal microscopic transient, where $\kappa_2\propto t^2$ and 
thus $W\propto t$.
This crosses over to the FV growth regime ($1\ll Jt\ll \ell$), characterized by the scaling form $W^{(H)}(\ell,t)\sim \ell^{\alpha_H} f_H(Jt/\ell)$ and the power law $W\propto t^{\beta_H}$.
Finally, for $t\gtrsim \ell/J$ the dynamics loses memory of the initial segment occupation and the roughness saturates to the FV steady value $W_{\rm sat}^{(H)}(\ell)\propto \ell^{\alpha_H}$.

\subsection{Lindbladian evolution and dephasing effects}
\label{sm:sec:lindblad_mu2_XX}

We now derive an analytical expression for the second moment under the \emph{full} Lindblad evolution in the XX limit, i.e., for $\Delta=0$ with homogeneous gain and loss processes at every site as in Sec.~\ref{sm:sec:QGF} and Eq.~\eqref{sm:eq:Dissipation}.
As in the unitary discussion, it is convenient to work with the fermion number in the segment, $N_\ell=\sum_{j\in\mathrm{seg}(\ell)} n_j$, so that $\Gamma(t)=N_\ell(t)-N_\ell(0)$, according to~Eq.~\eqref{sm:eq:Gamma_is_number_transfer}.
\begin{figure}[tbh!]
	\begin{center}
	 \includegraphics[width=0.4\columnwidth]{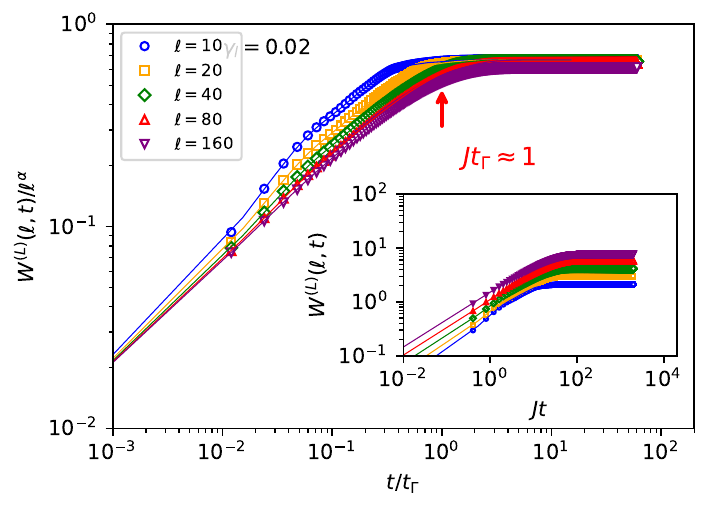}
	 \includegraphics[width=0.4\columnwidth]{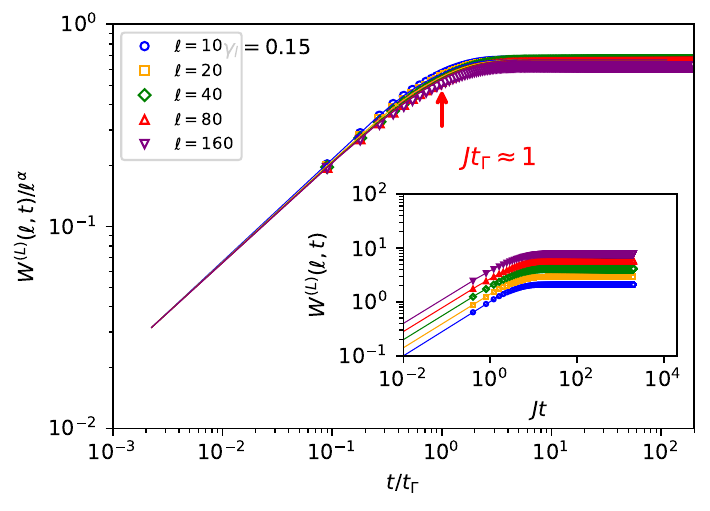}
	 \caption[Scaling collapse, dissipation-dominated]{Scaling collapse in the dissipation-dominated regime $t_{\Gamma}\sim \Gamma^{-1}\ll t^*$, where the roughness saturates already at $t\sim t_{\Gamma}$ while the ballistic scaling 
	 variable $x=Jt/\ell$ remains small.  The crossover time $t_{\Gamma}$ is indicated by the vertical arrow. The solid 
	 lines are the 
	 analytic result from Eq.~\eqref{sm:eq:mu2_lindblad_bessel}  while the symbols show the numerical TEBD 
	 data for $W^{(\mathcal{L})}(\ell,t)$ for the same set of parameters. System size is fixed to $L=500$  and $\zeta = 0.5$. 
	 The left panel corresponds to $\Gamma=0.02J$, while the right panel corresponds to $\Gamma=0.15J$. The vertical marker at $t_{\Gamma}$ highlights the crossover from the initial growth regime to saturation, which occurs at a time scale set by the dissipation rate rather than by the segment length. The solid lines show the analytic result from Eq.~\eqref{sm:eq:mu2_lindblad_bessel}, which captures the full time dependence of the roughness and agrees well with the numerical data.
	 The insets represent the unrescaled data, indicating that the three regimes clearly visible in the unitary evolution are no longer present, as the short time transient is now followed by a direct crossover to saturation at $t\sim t_{\Gamma}$. System sizes are $L=500$ and segment lengths $\ell=10, 20,40,80, 160$ and in the TEBD simulations we used a maximum bond dimension $\chi_{\rm max}=32$ and a time step $dt=0.01/J$.
	 }
	 \label{sm:fig:FV_Lindblad_Delta_0}
	\end{center}
\end{figure}
Because we evaluate dynamical fluctuations on top of the nonequilibrium steady state $\rho_{\mathrm{SS}}$ of the \emph{same} Liouvillian, the process is stationary and the equal-time variance is time independent,
\begin{equation}
\langle N_\ell(t)^2\rangle_{\mathrm{SS}}=\langle N_\ell^2\rangle_{\mathrm{SS}}.
\end{equation}
Hence the second moment can again be written as
\begin{align}
\mu_2^{(\mathcal{L})}(\ell,t)
&=\Big\langle\big[N_\ell(t)-N_\ell(0)\big]^2\Big\rangle_{\mathrm{SS}}\nonumber\\
&=2\Big(\langle N_\ell^2\rangle_{\mathrm{SS}}-\langle N_\ell(t)N_\ell\rangle_{\mathrm{SS}}\Big).
\label{sm:eq:mu2_lindblad_basic}
\end{align}
In the XX limit ($\Delta=0$), the Lindblad problem with homogeneous gain/loss is quadratic in Majorana fermions and can 
be solved exactly (e.g., within the third-quantization framework~\cite{Prosen.2008,Prosen2010}).
For bulk observables in a translation-invariant geometry (or for segments sufficiently far from the boundaries), the 
coherent part generates the same ballistic Bessel-function kernel as in Eq.~\eqref{sm:eq:U_bessel}, while the local 
incoherent spin flips provide a finite lifetime for phase-coherent propagation.
Equivalently, the reservoirs act as an effective dephasing/relaxation mechanism for dynamical \emph{fluctuations} on top 
of $\rho_{\mathrm{SS}}$, which damps the unitary propagation kernel by an exponential factor.
Denoting the corresponding decay rate by
\begin{equation}
\Gamma\equiv \gamma_l+\gamma_p,
\end{equation}
the connected density correlator takes the form
\begin{equation}
\langle n_i(t)n_j\rangle_{\mathrm{SS}}-\bar n^2
=\bar n(1-\bar n)\,e^{-\Gamma t}\,|U_{ij}(t)|^2,
\label{sm:eq:nn_lindblad_kernel}
\end{equation}
with $U_{ij}(t)$ given by the same Bessel-function kernel as in the unitary case, Eq.~\eqref{sm:eq:U_bessel}.
Substituting Eq.~\eqref{sm:eq:nn_lindblad_kernel} into Eq.~\eqref{sm:eq:mu2_lindblad_basic} yields a closed expression analogous to Eq.~\eqref{sm:eq:mu2_unitary_bessel},
\begin{align}
\mu_2^{(\mathcal{L})}(\ell,t)
&=2\bar n(1-\bar n)\left[\ell-e^{-\Gamma t}\sum_{r=-(\ell-1)}^{\ell-1}(\ell-|r|)\,J_r^2(Jt)\right].
\label{sm:eq:mu2_lindblad_bessel}
\end{align}

Equation~\eqref{sm:eq:mu2_lindblad_bessel} makes explicit how the dissipative couplings 
preserve the ballistic Bessel-function light-cone structure inherited from the XX Hamiltonian, 
while exponentially suppressing the two-time correlations that encode memory of the initial 
segment configuration.
In particular, for $t\to\infty$ the correlator $\langle N_\ell(t)N_\ell\rangle_{\mathrm{SS}}$ 
decorrelates and the second moment saturates to twice the static variance,
 $\mu_2^{(\mathcal{L})}(\ell,t\to\infty)\to 2\bar n(1-\bar n)\ell$, whereas for short 
 times the unitary result is recovered up to $\mathcal{O}(\Gamma t)$ corrections.

It is instructive to translate Eq.~\eqref{sm:eq:mu2_lindblad_bessel} into asymptotic forms for the roughness
$W^{(\mathcal{L})}(\ell,t)=\sqrt{\mu_2^{(\mathcal{L})}(\ell,t)}$.
In the pre-saturation regime where the dynamics is still dominated by coherent ballistic spreading,
\begin{equation}
1\ll Jt\ll \ell,\qquad \Gamma t\ll 1,
\end{equation}
one may use the same Bessel-weight asymptotics as in the unitary case,
$\sum_{r=-\infty}^{\infty}|r|J_r^2(Jt)\simeq \tfrac{2}{\pi}Jt$, to obtain
\begin{equation}
W^{(\mathcal{L})}(\ell,t)\simeq \left(\frac{4}{\pi}\bar n(1-\bar n)\,Jt\right)^{\!1/2},
\phantom{a} 1\ll Jt\ll \ell,\ \Gamma t\ll 1,
\label{sm:eq:W_lindblad_growth}
\end{equation}
which coincides with the unitary growth law.
In contrast, at long times the exponential damping suppresses the memory term in Eq.~\eqref{sm:eq:mu2_lindblad_bessel} and yields saturation to twice the static variance,
\begin{equation}
W^{(\mathcal{L})}_{\rm sat}(\ell)\simeq \big(2\bar n(1-\bar n)\big)^{\!1/2}\,\ell^{1/2},
\qquad t\gg \max\{\ell/J,\,\Gamma^{-1}\}.
\label{sm:eq:W_lindblad_sat}
\end{equation}
Accordingly, the natural FV scaling variable remains $x=Jt/\ell$ (ballistic $z=1$), but dissipation introduces a second dimensionless time parameter $y\equiv \Gamma t$.
Defining $f_{\mathcal{L}}(x,y)$ through $W^{(\mathcal{L})}(\ell,t)=\ell^{1/2} f_{\mathcal{L}}(x,y)$, Eqs.~\eqref{sm:eq:W_lindblad_growth}--\eqref{sm:eq:W_lindblad_sat} imply the asymptotics
\begin{equation}
f_{\mathcal{L}}(x,y)\simeq
\begin{cases}
\left(\frac{4}{\pi}\bar n(1-\bar n)\,x\right)^{\!1/2}, & x\ll 1\ \text{and}\ y\ll 1,\\
\big(2\bar n(1-\bar n)\big)^{\!1/2}, & x\gg 1\ \text{or}\ y\gg 1.
\end{cases}
\label{sm:eq:fL_asymptotics}
\end{equation}

Once the dissipative time scale becomes relevant, $y=\Gamma t\gtrsim 1$, the exponential factor in Eq.~\eqref{sm:eq:mu2_lindblad_bessel} suppresses the ``memory'' term
$\sum_r(\ell-|r|)J_r^2(Jt)$ and drives the system toward the plateau
$W^{(\mathcal{L})}(\ell,t)\to \sqrt{2\bar n(1-\bar n)\,\ell}$.
This introduces a second crossover time scale, $t_{\Gamma}\sim \Gamma^{-1}$, in addition to the coherent finite-size time $t^*\sim \ell/J$.
The unitary-like FV growth regime therefore requires
\begin{equation}
t\ll \min\{t^*,\,t_{\Gamma}\}=\min\{\ell/J,\,\Gamma^{-1}\}.
\end{equation}
In the weak-dissipation limit $t_{\Gamma}\gg t^*$ (equivalently $\Gamma\ll J/\ell$), one observes the same ballistic scaling with $x=Jt/\ell$ as in the unitary case.
In the opposite, \emph{dissipation-dominated} regime $t_{\Gamma}\ll t^*$ (i.e., $\Gamma\gg J/\ell$), the roughness saturates already at $t\sim t_{\Gamma}$ while the ballistic scaling variable is still small, $x\sim J/(\Gamma\ell)\ll 1$.
In that case a one-parameter collapse in $x$ alone is not expected beyond very short times, and the appropriate description is the two-parameter scaling function $f_{\mathcal{L}}(x,y)$ with $y$ controlling the approach to saturation.

In this limit, the FV collapse in terms of the single ballistic variable $x$ breaks down, since saturation is controlled by $t_{\Gamma}$ rather than by $t^*$.
One may nevertheless view the approach to the plateau at fixed $x\ll 1$ as a crossover governed primarily by $y=\Gamma t$.
For example, expanding Eq.~\eqref{sm:eq:mu2_lindblad_bessel} at short times gives the leading behavior
$W^{(\mathcal{L})}(\ell,t)\propto \ell^{1/2}(\Gamma t)^{1/2}$ when the dissipative term dominates, while for $y\gg 1$ the roughness saturates to $W_{\rm sat}^{(\mathcal{L})}(\ell)$.

In Fig.~\ref{sm:fig:FV_Lindblad_Delta_0} we show the scaling collapse for the Lindbladian evolution in the XX limit with 
$\gamma_l=0.02 J$ and $\zeta=0.5$, which corresponds to the dissipation-dominated regime $t_{\Gamma}\sim \Gamma^{-1}\ll t^*$ for the system sizes considered. The symbols show the numerical TEBD data for $W^{(\mathcal{L})}(\ell,t)$ while the 
solid lines show the analytic result from Eq.~\eqref{sm:eq:mu2_lindblad_bessel} for the same parameters. The vertical arrow 
indicates the crossover time $t_{\Gamma}$, which controls the approach to saturation in this regime. 

For small dissipation rates $\Gamma$ there appears to be a  competition between the \emph{coherent} FV scaling variable $x=Jt/\ell$ (which organizes the 
unitary growth-and-saturation crossover at $t\sim t^*$) and the \emph{dissipative} scaling variable $y=\Gamma t$ (which encodes the loss of memory and drives 
relaxation at $t\sim t_{\Gamma}$). As $\Gamma$ is increased, the separation of time scales is reduced and eventually inverted: $t_{\Gamma}=\Gamma^{-1}$ 
becomes parametrically shorter than $t^*\sim \ell/J$, so the dynamics reaches its plateau before the subsystem explores the coherent finite-size window. In this regime, the approach to saturation is controlled predominantly by $t/t_{\Gamma}$ (or $y$), and $t_{\Gamma}$ becomes the only relevant time scale for the 
scaling collapse over the accessible time window.

Notice that for this particular loss rate $\gamma_l=0.02 J$, the system is in the dissipation-dominated regime for all segment lengths $\ell\gtrsim 70$, so that the collapse in terms of the ballistic variable $x=Jt/\ell$ breaks down beyond very short times, as highlighted in the inset of the upper panel in Fig.~\ref{sm:fig:FV_Lindblad_Delta_0}. In contrast, for larger dissipation $\gamma_l=0.15 J$ the system is in the dissipation-dominated regime for all segment lengths considered, so that the collapse is controlled by $t_\Gamma$, as shown in the lower panel of Fig.~\ref{sm:fig:FV_Lindblad_Delta_0}.

Physically, this failure of the standard (unitary-like) FV scaling collapse at strong dissipation reflects the fact that the reservoirs impose a finite coherence/relaxation time $t_{\Gamma}\sim \Gamma^{-1}$ for dynamical fluctuations.
When $t_{\Gamma}\ll t^*\sim \ell/J$, the system reaches its fluctuation plateau before the subsystem can explore the ballistic finite-size window, so the roughness is no longer controlled by the ratio $t/\ell$ but by the dissipation.
As a result,  a one-parameter FV collapse is not expected except in the earliest transient; the appropriate description is instead the two-parameter scaling form $W^{(\mathcal{L})}(\ell,t)=\ell^{\alpha} f_{\mathcal{L}}(Jt/\ell,\Gamma t)$ (or, in the dissipation-dominated regime, an effective collapse primarily in $y$ at fixed $x\ll 1$).

\section[Interacting case Delta!=0]{Interacting case \texorpdfstring{$\Delta\neq 0$}{Delta!=0}}
\label{sm:sec:interacting}

\subsection{Unitary evolution}\label{sm:sec:unitary_Delta_nonzero}
\begin{figure}[tbh!]
	\begin{center}
	 \includegraphics[width=0.4\columnwidth]{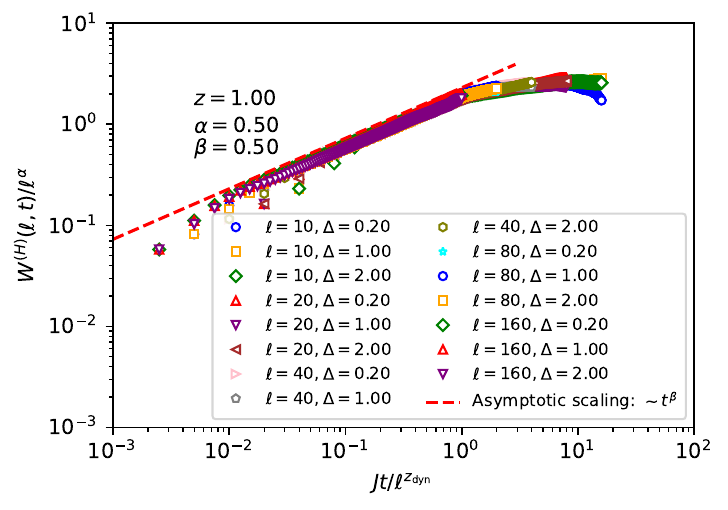}
	 \caption{ FV scaling collapse for the unitary evolution in the interacting XXZ 
	 chain with $\Delta=\{0.2, 1.0, 2.0\}$ and $\zeta=0.1$. The symbols show the numerical TEBD data while the 
	 dashed  lines are the asymptotic forms of the universal function corresponding to the growth $\sim t^{\beta}$.
	 System sizes are $L=500$ and segment lengths $\ell=10, 20,40,80, 160$ and in the TEBD simulations we use maximum 
	 bond dimension $\chi_{\rm max}=32$ and a time step $dt=0.01/J$.}
	 \label{sm:fig:FV_unitary_z_0.1}
	\end{center}
\end{figure}

\begin{figure*}[tbh!]
	\begin{center}
	 \includegraphics[width=0.32\columnwidth]{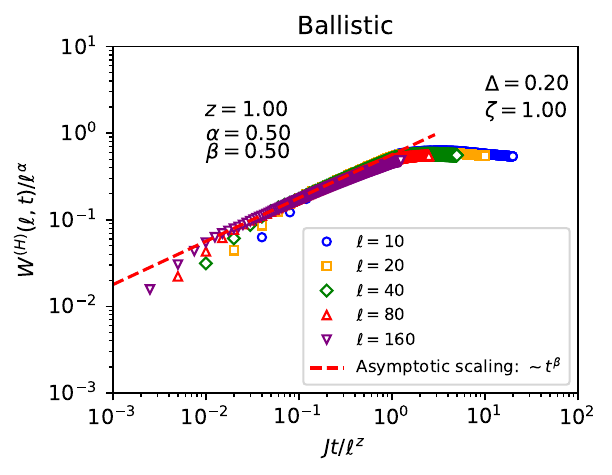}
	 \includegraphics[width=0.32\columnwidth]{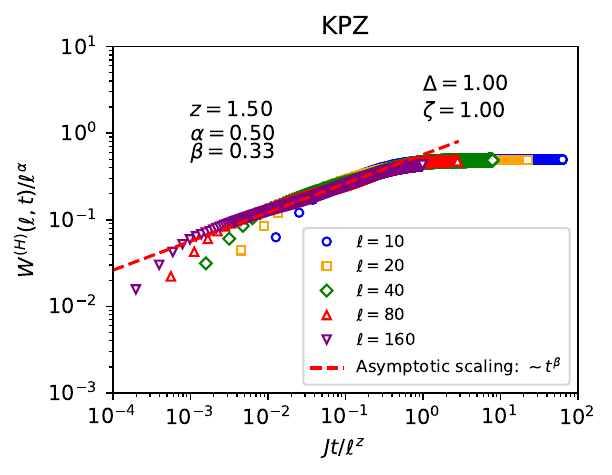}
	 \includegraphics[width=0.3\columnwidth]{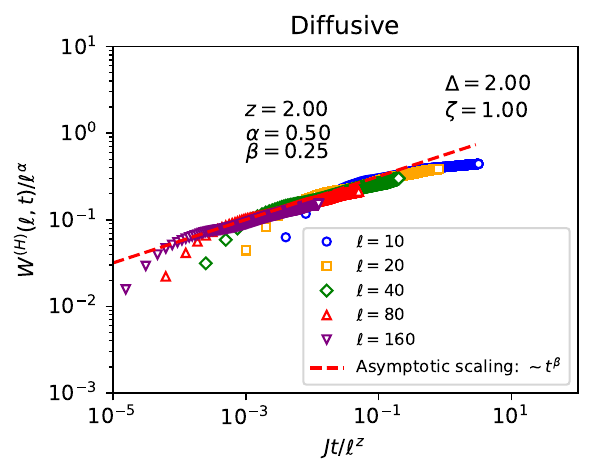}	
	 \caption{Scaling collapse for the unitary evolution in the 
	 interacting regime, at infinite temperature $\zeta=1.0$ and $\Delta=\{0.2, 1.0, 2.0\}$.
	 For $\Delta<1$, the scaling collapse is compatible with the same ballistic scaling variable $x=Jt/\ell$ as in the unitary case, while for $\Delta\geq 1$ the collapse is achieved with a diffusive scaling variable $x=Jt/\ell^2$.
	 At $\Delta=1$, the system is at the isotropic point, where the scaling behavior is compatible with the KPZ universality class, with $x=Jt/\ell^{3/2}$.
	 The symbols show the numerical TEBD data for $W^{(\mathcal{L})}(\ell,t)$ while the dashed lines 
	 indicate the expected short-time growth $\sim (\Gamma t)^{1/2}$. System sizes are $L=500$ and segment lengths 
	 $\ell=10, 20,40,80,160$; TEBD parameters are $\chi_{\rm max}=32$ and $dt=0.01/J$.}
	 \label{sm:fig:FV_Unitary_Delta_1}
	\end{center}
\end{figure*}

In the interacting XXZ chain, closed analytic expressions for the time-dependent QGF and the roughness $W^{(H)}(\ell,t)$ 
are not available.
We therefore extract FV scaling from numerical time evolution using the same protocol as in the noninteracting case: the 
chain is prepared in the diagonal product NESS $\rho_{\mathrm{SS}}(\zeta)$ of the driven--dissipative problem and then 
evolved unitarily with the interacting Hamiltonian $H$.
The roughness is obtained from the second cumulant of the transferred segment magnetization, $W^{(H)}(\ell,t)=\sqrt{\kappa_2^{(H)}(\ell,t)}$, evaluated via our vectorized TEBD implementation of the QGF (Sec.~\ref{sm:sec:QGF}).
As before, the saturation with subsystem size is consistent with
\begin{equation}
W^{(H)}_{\rm sat}(\ell)\propto \ell^{\alpha},\qquad \alpha\simeq \frac{1}{2},
\end{equation}
reflecting the essentially uncorrelated nature of the initial product state and the extensive character of the segment 
observable.
The dynamical information is therefore primarily encoded in the crossover time scale and in the growth regime at fixed $\ell$.

Figure~\ref{sm:fig:FV_unitary_z_0.1} shows that for finite magnetization density ($\zeta\neq 1$) the rescaled roughness curves obtained at different anisotropies $\Delta$ collapse onto the same universal curve.
In other words, once $W^{(H)}(\ell,t)$ is expressed in FV form and plotted against the single scaling variable $t/\ell$, the interaction strength (here tuned by $\Delta$) primarily renormalizes nonuniversal amplitudes and the microscopic transient, but does not spoil the overall collapse.
This robust data collapse across $\Delta=\{0.2,1.0,2.0\}$ provides strong evidence that the finite-density unitary dynamics belongs to a common ballistic scaling scenario for the FV roughness.
For generic fillings $\zeta\neq 1$ (finite magnetization density), the FV scaling collapse extracted from TEBD is compatible with a ballistic universality class,
\begin{equation}
W^{(H)}(\ell,t)\sim \ell^{1/2}\, f\!\left(\frac{t}{\ell}\right),\qquad z\simeq 1,
\end{equation}
in the sense that the collapse is achieved with a single scaling variable $t/\ell$ and the 
crossover occurs at times $t^*\propto \ell$.
This behavior can be understood qualitatively from the fact that away from half filling the interacting XXZ chain 
supports stable quasiparticle propagation (magnons), leading to finite-velocity 
transport and thus $z=1$~\cite{Schmitteckert.2020}.
In particular, Ref.~\cite{Schmitteckert.2020} analyzes high-temperature spin dynamics in the (integrable) Heisenberg/XXZ 
chain and shows that, at finite magnetization density, spin correlations are largely carried by long-lived magnon-like 
quasiparticles that generate a  ballistic ``light-cone'' front.
The front position is set by the quasiparticle group velocity (hence a characteristic $t\propto \ell$ crossover), while 
dissipationless many-body scattering manifests mainly in the broadening and shape of the propagating signal rather than 
in a change of the ballistic dynamical expone, a scenario that is consistent with the observed FV collapse in Fig.~\ref{sm:fig:FV_unitary_z_0.1}. As indicated by the 
dashed lines in Fig.~\ref{sm:fig:FV_unitary_z_0.1}, the growth regime is compatible with
\begin{equation}
W^{(H)}(\ell,t)\simeq \sqrt{\frac{4}{\pi}\bar n(1-\bar n)}\left( \,Jt\right)^{\beta},\qquad 1\ll Jt\ll \ell,
\end{equation}
The infinite-temperature case $\zeta=1$ is special because the initial state has vanishing magnetization. 
This regime was investigate before in Ref.~\cite{Moca.2026}. 
For completeness, we also show the FV collapse for $\zeta=1$ in Fig.~\ref{sm:fig:FV_Unitary_Delta_1}.
In this case, the extracted dynamical exponent depends on the anisotropy $\Delta$, 
in line with the known variety of transport regimes in the XXZ chain.
In the easy-plane regime ($|\Delta|<1$) the collapse is consistent with ballistic scaling, $z\simeq 1$.
At the isotropic point ($\Delta=1$), the growth and crossover are compatible with 
superdiffusive KPZ-type scaling, $z\simeq 3/2$.
In the easy-axis regime ($\Delta>1$), the scaling crosses over to diffusive behavior with $z\simeq 2$.
This phenomenology is consistent with the known transport properties of the XXZ chain at infinite temperature, where the spin transport is ballistic for $|\Delta|<1$, superdiffusive at $\Delta=1$, and diffusive for $\Delta>1$~\cite{ Ljubotina.2017, Ljubotina_2017}.

\subsection{Lindbladian evolution}\label{sm:sec:lindbladian_Delta_nonzero}
\begin{figure}[tbh!]
	\begin{center}
	 \includegraphics[width=0.4\columnwidth]{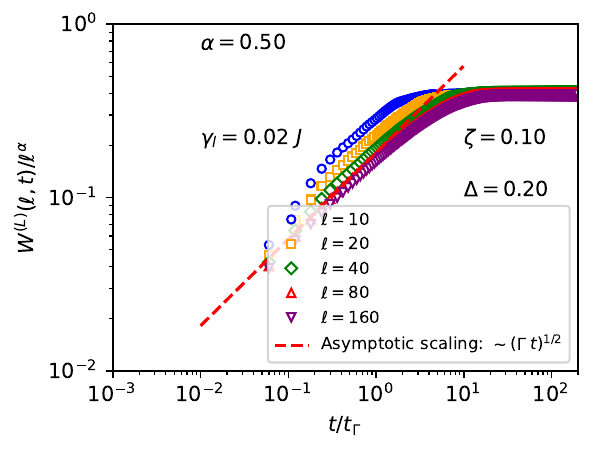}
	 \includegraphics[width=0.4\columnwidth]{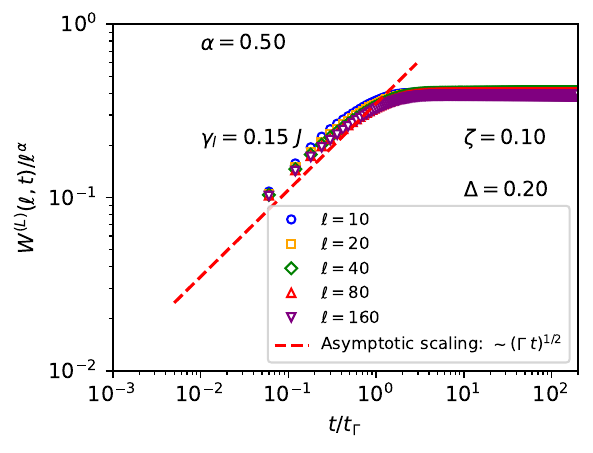}
	 \caption[Breakdown of FV collapse under dissipation]{Breakdown of the scaling collapse for the Lindbladian evolution in the interacting regime, highlighting the breakdown of a one-parameter Family--Vicsek collapse in terms of the coherent 
	 scaling variable $Jt/\ell$ and the crossover to a dissipation-controlled saturation governed by $t_{\Gamma}\sim \Gamma^{-1}$ with increasing $\gamma_l$. The left panel corresponds to $\Delta=0.2$, $\zeta=0.1$ and $\gamma_l=0.02J$, while the right panel corresponds to  $\gamma_l=0.15J$. In both cases, irrespective of the value of $\Delta$, and $\zeta$, the approach to saturation is controlled by the dissipative time scale $t_{\Gamma}\sim \Gamma^{-1}$.
	 The symbols show the numerical TEBD data for $W^{(\mathcal{L})}(\ell,t)$ while the dashed lines 
	 indicate the expected short-time growth $\sim (\Gamma t)^{1/2}$. System sizes are $L=500$ and segment lengths 
	 $\ell=10, 20,40,80,160$; TEBD parameters are $\chi_{\rm max}=32$ and $dt=0.01/J$.}
	 \label{sm:fig:FV_lindbladian_Delta_0.2}
	\end{center}
\end{figure}
In the interacting case, turning on bulk gain and loss qualitatively changes the scaling structure of the roughness dynamics.
While the unitary evolution admits a one-parameter FV collapse with a transport-controlled crossover time $t^*\propto \ell^{z}$ (Sec.~\ref{sm:sec:unitary_Delta_nonzero}), the full Lindblad evolution introduces an additional, \emph{microscopic} relaxation time scale set by the local jump rates.
For homogeneous gain/loss, the relevant dissipative rate is
\begin{equation}
\Gamma\equiv \gamma_l+\gamma_p,\qquad t_{\Gamma}\sim \Gamma^{-1},
\end{equation}
which controls the decay of two-time correlations (and therefore of the QGF) irrespective of the coherent interaction strength.

As a result, a collapse of $W^{(\mathcal{L})}(\ell,t)$ as a function of the single FV scaling variable $x=t/\ell^{z}$ generally \emph{fails} in the interacting driven--dissipative dynamics: for the system sizes and rates considered here, the approach to the plateau occurs already when $t\sim t_{\Gamma}$, i.e., before the coherent finite-size time $t^*$ is reached.
Equivalently, the roughness is better viewed as a two-parameter scaling form,
\begin{equation}
W^{(\mathcal{L})}(\ell,t)=\ell^{1/2}\,f_{\mathcal{L}}\!\left(\frac{Jt}{\ell},\,\Gamma t\right),
\end{equation}
where the second argument $y\equiv \Gamma t$ drives the loss of memory of the initial segment configuration.
In this dissipation-dominated regime, the data for different $\ell$ do not collapse when plotted against $Jt/\ell$ beyond very short times, because the relevant crossover is set primarily by $t_{\Gamma}$ rather than by $t^*$.

The same physics also implies a simple and robust scaling collapse when time is measured in units of $t_{\Gamma}$.
Indeed, for all fugacities $\zeta$ (including $\zeta=1$) and for all anisotropies $\Delta$ shown in Fig.~\ref{sm:fig:FV_lindbladian_Delta_0.2}, the curves are consistent with a relaxation scenario in which
\begin{equation}
W^{(\mathcal{L})}(\ell,t)\simeq \ell^{1/2}\,g(\Gamma t),
\end{equation}
with a short-time growth compatible with $g(y\ll 1)\propto y^{1/2}$ and a saturation to a constant for $y\gg 1$.
Thus, interactions and filling primarily affect the very early-time coherent transient, while the breakdown of 
one-parameter FV scaling and the collapse with $t_{\Gamma}$ are controlled by the local dissipative processes. 
Furthermore, for $\zeta=0.1$ and $\gamma_l = 0.02$ the curves collapse can be estimated for segment lengths larger than $\ell_c= \pi J/2\Gamma\simeq 71$ while for larger
$\gamma_l=0.15$ the collapse is controlled by $t_{\Gamma}$ for all segment lengths considered, as indicated by the lower panel of Fig.~\ref{sm:fig:FV_lindbladian_Delta_0.2}.
\section{Integrability-breaking perturbations}
\label{sm:sec:integrability_breaking}

The robustness of the FV scaling collapse under integrability-breaking perturbations is an interesting open question.
In this section we address this question and consider the effect of next-nearest neighbor interactions, which break the integrability of the XXZ chain but preserve the $U(1)$ symmetry and the product form of the NESS. Specifically, we consider the Hamiltonian introduced in Eq.~\eqref{sm:eq:H_nnn} and the same bulk gain and loss processes as discussed in  Sec.~\ref{sm:sec:model_ness}.

\subsection{Noninteracting limit \texorpdfstring{$\Delta=0$}{Delta=0}: unitary evolution with \texorpdfstring{$J_2\neq 0$}{J2!=0}}
\label{sm:sec:unitary_nnn_Delta_0}
Before turning to the interacting integrability-breaking dynamics ($\Delta\neq 0$), it is useful to consider the special case $\Delta=0$ with a finite next-nearest-neighbor coupling $J_2$, a limit which is not integrable breakingly but remains exactly solvable via free-fermion techniques.
While $H_{\rm nnn}$ breaks integrability once interactions are present, at $\Delta=0$ the model remains quadratic after the Jordan--Wigner transformation; the main effect is that the single-particle dispersion is modified compared to the nearest-neighbor XX chain.
Consequently, the exact free-fermion evaluation of the unitary roughness can be repeated with a modified propagation kernel, providing an analytic benchmark for the FV scaling structure.
At $\Delta=0$, the coherent Hamiltonian $H+H_{\rm nnn}$ maps to spinless fermions with nearest- and next-nearest-neighbor hopping, with single-particle dispersion
$\varepsilon(k)= -J\cos k - J_2\cos(2k)$,
and group velocity $v(k)=\partial_k\varepsilon(k)=J\sin k+2J_2\sin(2k)$.
Because the steady state $\rho_{\mathrm{SS}}$ is still a diagonal product state and commutes with the Hamiltonian (Sec.~\ref{sm:sec:model_ness}), the second moment for the unitary quench remains of the form in Eq.~\eqref{sm:eq:mu2_unitary_basic}.
Repeating the Wick-contraction steps summarized in Appendix~\ref{sm:app:unitary_mu2} yields the exact expression
as in Eq.~\eqref{sm:eq:mu2_unitary_bessel_app} but with the modified kernel $U_r(t)$ given by the Fourier transform of the dispersion,
\begin{equation}
U_r(t)=\int_{-\pi}^{\pi}\!\frac{dk}{2\pi}\,e^{-i\varepsilon(k)t}\,e^{ikr}.
\label{sm:eq:U_r_nnn}
\end{equation}
That implies that the FV exponents remain those of a ballistic growth-and-saturation scenario.
In particular, the saturation value is unchanged,
\begin{equation}
W^{(H)}_{\rm sat}(\ell)\simeq \big(2\bar n(1-\bar n)\big)^{\!1/2}\,\ell^{1/2},
\label{sm:eq:W_unitary_nnn_sat}
\end{equation}
because it is fixed by the equal-time variance of the initial product state.
In the pre-saturation window, a semiclassical quasiparticle estimate shows a  
ballistic spreading of the single-particle kernel, giving
\begin{equation}
W^{(H)}(\ell,t)\simeq \Big(2\bar n(1-\bar n)\,\bar v\,t\Big)^{\!1/2}, 1\ll t\,\max\{|J|,|J_2|\}\ll \ell/\bar v,
\label{sm:eq:W_unitary_nnn_growth}
\end{equation}
in terms of the effective velocity
\begin{equation}
\bar v\equiv \int_{-\pi}^{\pi}\!\frac{dk}{2\pi}\,|v(k)|=\int_{-\pi}^{\pi}\!\frac{dk}{2\pi}\,\big|J\sin k+2J_2\sin(2k)\big|.
\label{sm:eq:vbar_nnn}
\end{equation}
For example, for $J_2=J$, as considered in our numerics, 
one has $v(k)=J\sin k\,(1+4\cos k)$ and the integral can be evaluated in closed form,
\begin{equation}
\bar v\big|_{J_2=J}=\frac{17}{4\pi}\,J\simeq 1.35\,J,
\end{equation}
Matching Eqs.~\eqref{sm:eq:W_unitary_nnn_growth} and \eqref{sm:eq:W_unitary_nnn_sat} yields a crossover time scale $t^*\simeq \ell/\bar v$.
This indicates a change in the  FV crossover time which is reduced to 
\begin{equation}
t^*\simeq \ell/\bar v\simeq (4\pi/17)(\ell/J)\approx 0.74\,\ell/J.
\label{sm:eq:tstar_nnn}
\end{equation}
Thus, for $\Delta=0$ with $J_2\neq 0$ the FV scaling form
\begin{equation}
W^{(H)}(\ell,t)\sim \ell^{1/2}\, f_{H,\,\mathrm{nnn}}\!\left(\frac{\bar v t}{\ell}\right)
\label{sm:eq:FV_unitary_nnn}
\end{equation}
holds with $\alpha=\beta=1/2$ and $z=1$, and the scaling function has the universal asymptotics
\begin{equation}
	f_{H,\,\mathrm{nnn}}(x)\propto
	\begin{cases}
		 \big(2\bar n(1-\bar n)\big)^{\!1/2}\,x^{1/2},\qquad x\ll 1,\\
		 \big(2\bar n(1-\bar n)\big)^{\!1/2},\qquad x\gg 1.	
	\end{cases}
\end{equation}

\begin{figure}[tbh!]
	\begin{center}
	 \includegraphics[width=0.4\columnwidth]{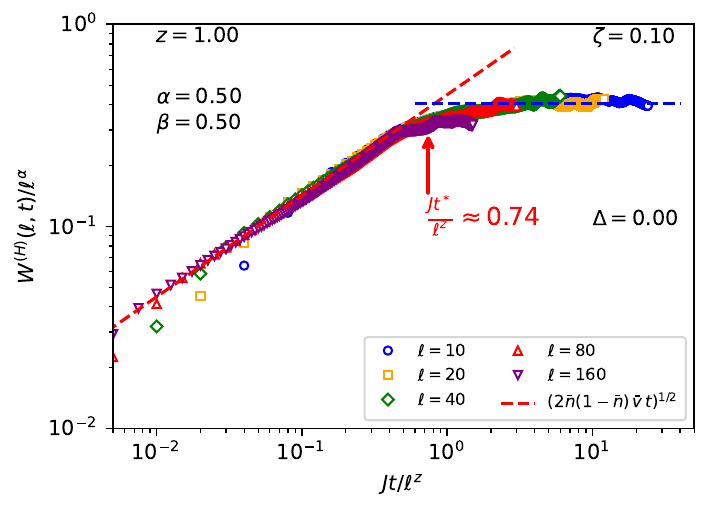}
	 \caption[FV scaling collapse, XX limit]{FV scaling collapse for the unitary evolution in the $\Delta=0$ limit but 
	 with a next-nearest neighbor coupling $J_2=J$ indicating a ballistic scaling scenario with a renormalized velocity $\bar v$ compared to the nearest-neighbor XX chain.  The symbols show the numerical TEBD data for 
	 $W^{(H)}(\ell,t)$.  The crossover time $t^*$ is estimated from Eq.~\eqref{sm:eq:tstar_nnn} and indicated by the vertical  
	 arrow. The universal function  is shown as dashed lines, with the growth and saturation regimes corresponding to 
	 Eqs.~\eqref{sm:eq:W_unitary_nnn_growth}   and \eqref{sm:eq:W_unitary_nnn_sat}. System sizes are $L=500$ and segment lengths $\ell=10, 20,40,80, 160$ and in the TEBD simulations we used a maximum bond dimension $\chi_{\rm max}=32$ and a time step $dt=0.01/J$.
	 }
	 \label{sm:fig:FV_nnn_unitary_Delta_0}
	\end{center}
\end{figure}

Figure~\ref{sm:fig:FV_nnn_unitary_Delta_0} illustrates that including $J_2\neq 0$ in the $\Delta=0$ unitary dynamics does not alter the FV exponents, but it does renormalize the ballistic time scale that controls the crossover to saturation.
When plotted in the FV variables, the TEBD data for different segment lengths $\ell$ collapse onto a single curve that exhibits the expected growth $W\propto t^{1/2}$ for $\bar v t/\ell\ll 1$ followed by saturation $W_{\rm sat}\propto \ell^{1/2}$ for $\bar v t/\ell\gg 1$.
The main quantitative effect of the next-nearest-neighbor hopping is therefore captured by replacing the nearest-neighbor velocity scale by the effective velocity $\bar v$ in Eq.~\eqref{sm:eq:vbar_nnn}, consistent with the shift of the crossover time $t^*$ indicated in the figure.
\subsection{Interacting case \texorpdfstring{$\Delta\neq 0$}{Delta!=0}: Unitary evolution with \texorpdfstring{$J_2\neq 0$}{J2!=0}}
\begin{figure}[tbh!]
	\begin{center}
	 \includegraphics[width=0.4\columnwidth]{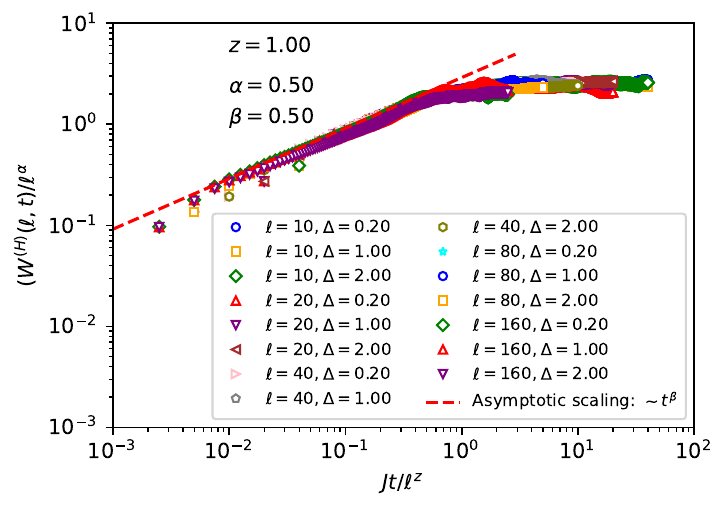}
	 \caption{ FV scaling collapse for the unitary evolution in the interacting non-intergrable XXZ chain with next-nearest neighbor interactions,
	  with $\Delta=\{0.2, 1.0, 2.0\}$ and $\zeta=0.1$. The symbols show the numerical TEBD data while the 
	 dashed  lines are the asymptotic forms of the universal function corresponding to the growth $\sim t^{\beta}$.
	 System sizes are $L=500$ and segment lengths $\ell=10, 20,40,80, 160$ and in the TEBD simulations we use maximum 
	 bond dimension $\chi_{\rm max}=32$ and a time step $dt=0.01/J$.}
	 \label{sm:fig:FV_nnn_unitary_z_0.1}
	\end{center}
\end{figure}

Figure~\ref{sm:fig:FV_nnn_unitary_z_0.1} shows that, at finite fugacity $\zeta\neq 1$, the ballistic unitary scaling phenomenology of the integrable XXZ chain remains visible even after integrability is broken by $J_2\neq 0$.
The observed collapse for different segment lengths $\ell$ is consistent with a quasiparticle (magnon) picture in which magnetization is carried by coherently propagating excitations, so that the dominant FV time scale is still set by a ballistic traversal time $t^*\propto \ell$.
Compared to the integrable case, the main effect of the integrability-breaking perturbation is to renormalize the effective propagation velocity (and, more generally, the distribution of velocities), which shifts the location of the crossover without destroying the collapse in the explored time window.

This behavior is in contrast to the infinite-temperature point $\zeta=1$, where integrability breaking leads to a qualitatively different transport scenario: the unitary dynamics becomes effectively diffusive, with a dynamical exponent compatible with $z=2$ irrespective of the value of $\Delta$. We have explored this $\zeta=1$ limit in detail in our previous work~\cite{Moca.2026}, and we shall not discuss it further here. 

\subsection{Interacting case \texorpdfstring{$\Delta\neq 0$}{Delta!=0}: Lindbladian evolution with \texorpdfstring{$J_2\neq 0$}{J2!=0}}
\label{sm:sec:lindbladian_nnn_Delta_nonzero}
\begin{figure}[tbh!]
	\begin{center}
	 \includegraphics[width=0.4\columnwidth]{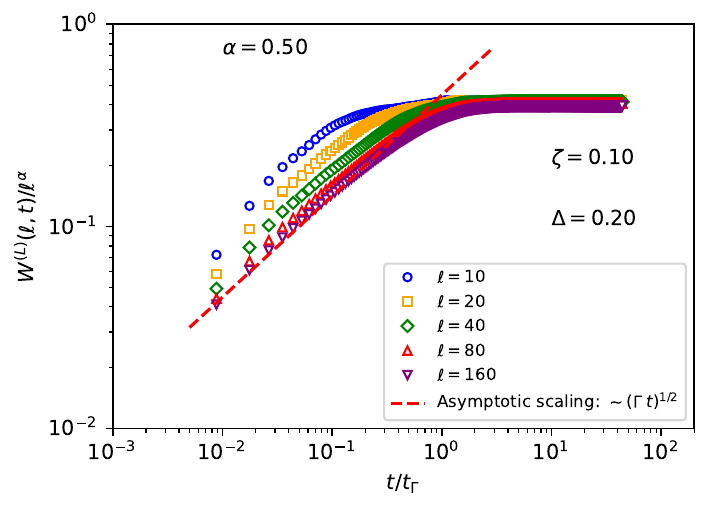}
	 \includegraphics[width=0.4\columnwidth]{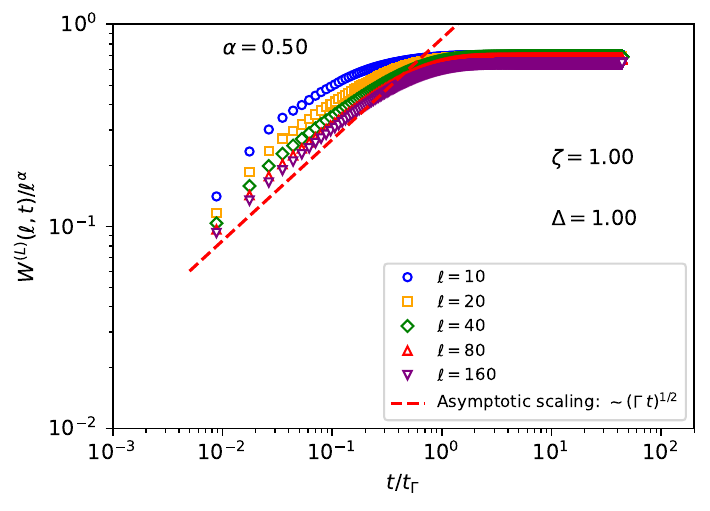}
	 \caption[Breakdown of FV collapse under dissipation]{Breakdown of the scaling collapse for the Lindbladian evolution in the non-integrable XXZ chain with finite $J_2$, highlighting the breakdown of a one-parameter Family--Vicsek collapse. The left panel corresponds to $\Delta=0.2$ and $\zeta=0.1$, while the right panel corresponds to $\Delta=1.0$ and $\zeta=1.0$. In both cases, irrespective of the value of $\Delta$, and $\zeta$, the approach to saturation is controlled by the dissipative time scale $t_{\Gamma}\sim \Gamma^{-1}$.
	 The symbols show the numerical TEBD data for $W^{(\mathcal{L})}(\ell,t)$ while the dashed lines 
	 indicate the expected short-time growth $\sim (\Gamma t)^{1/2}$. System sizes are $L=500$ and segment lengths 
	 $\ell=10, 20,40,80,160$; TEBD parameters are $\chi_{\rm max}=32$ and $dt=0.01/J$. We use $\gamma_l=0.02J$ and $\zeta=0.1,1.0$ for the left and right panels, respectively.}
	 \label{sm:fig:FV_nnn_lindbladian_Delta_1}
	\end{center}
\end{figure}
Figure~\ref{sm:fig:FV_nnn_lindbladian_Delta_1} demonstrates that the breakdown of one-parameter FV scaling under Lindbladian evolution persists in the presence of integrability-breaking next-nearest-neighbor interactions.
For both $\zeta\neq 1$ and $\zeta=1$ (and for the representative anisotropies shown), the roughness approaches its plateau on the dissipative time scale $t_{\Gamma}\sim \Gamma^{-1}$, so that plotting against the coherent scaling variable $Jt/\ell$ does not yield a sustained collapse beyond very short times.
Thus, non-integrability ($J_2\neq 0$) does not qualitatively change the Lindbladian scaling scenario compared to the integrable case shown in Fig.~\ref{sm:fig:FV_lindbladian_Delta_0.2}: dissipation controls the crossover and dominates the loss of two-time memory required for the QGF-based FV analysis.

\appendix
\section{Unitary evolution in the XX limit}
\label{sm:app:unitary_mu2}

This Appendix provides details on evaluating Eq.~\eqref{sm:eq:mu2_unitary_basic}.  First, we prove the identity
\begin{equation}
\langle N_\ell(t)^2\rangle_{\mathrm{SS}}=\langle N_\ell^2\rangle_{\mathrm{SS}}.
\label{sm:eq:proof_stationarity_Nl2}
\end{equation}
Under unitary evolution, $N_\ell(t)=e^{iHt}N_\ell e^{-iHt}$ and therefore $N_\ell(t)^2=e^{iHt}N_\ell^2 e^{-iHt}$.
Using cyclicity of the trace,
\begin{align}
\langle N_\ell(t)^2\rangle_{\mathrm{SS}}
&\equiv \mathrm{Tr}\!\left[\rho_{\mathrm{SS}}\,e^{iHt}N_\ell^2 e^{-iHt}\right]\nonumber\\
&=\mathrm{Tr}\!\left[e^{-iHt}\rho_{\mathrm{SS}}\,e^{iHt}N_\ell^2\right].
\end{align}
Since the steady state is diagonal in the $\sigma^z$ basis it can be written as a function of the conserved charge $S^z$, e.g.,
$\rho_{\mathrm{SS}}\propto e^{\mu S^z}$ (equivalently $\rho_{\mathrm{SS}}\propto \zeta^{S^z+L/2}$), and because $[H,S^z]=0$ we have $[H,\rho_{\mathrm{SS}}]=0$.
Hence $e^{-iHt}\rho_{\mathrm{SS}}e^{iHt}=\rho_{\mathrm{SS}}$, which directly implies Eq.~\eqref{sm:eq:proof_stationarity_Nl2}.
Using Eq.~\eqref{sm:eq:Gamma_is_number_transfer} and expanding the square,
\begin{align}
\mu_2^{(H)}(\ell,t)&=\left\langle\big[N_\ell(t)-N_\ell(0)\big]^2\right\rangle_{\mathrm{SS}}\nonumber\\
&=\langle N_\ell(t)^2\rangle_{\mathrm{SS}}+\langle N_\ell^2\rangle_{\mathrm{SS}}-2\langle N_\ell(t)N_\ell\rangle_{\mathrm{SS}}.
\end{align}
Using the stationarity proved above, $\langle N_\ell(t)^2\rangle_{\mathrm{SS}}=\langle N_\ell^2\rangle_{\mathrm{SS}}$, yields Eq.~\eqref{sm:eq:mu2_unitary_basic}. For the product steady state with $\langle n_j\rangle_{\mathrm{SS}}=\bar n$ and $n_j^2=n_j$,
\begin{align}
\langle N_\ell^2\rangle_{\mathrm{SS}}
&=\sum_{i,j\in\mathrm{seg}(\ell)}\langle n_i n_j\rangle_{\mathrm{SS}}\nonumber\\
&=\sum_{i\in\mathrm{seg}(\ell)}\langle n_i\rangle_{\mathrm{SS}}+\sum_{i\neq j\in\mathrm{seg}(\ell)}\langle n_i\rangle_{\mathrm{SS}}\langle n_j\rangle_{\mathrm{SS}}\nonumber\\
&=\ell\bar n+\ell(\ell-1)\bar n^2,
\label{sm:eq:Nl2}
\end{align}
which is Eq.~\eqref{sm:eq:Nl2}.
In the XX limit, the evolution is linear in fermions, according to Eq.~\eqref{sm:eq:cj_evolution}, and therefore
\begin{equation}
n_i(t)=c_i^\dagger(t)c_i(t)=\sum_{m,n}U_{im}^*(t)U_{in}(t)\,c_m^\dagger c_n.
\end{equation}
For the Gaussian steady state, the only non-vanishing two-point contractions are
\begin{equation}
\langle c_m^\dagger c_n\rangle_{\mathrm{SS}}=\bar n\,\delta_{mn},\qquad
\langle c_m c_n^\dagger\rangle_{\mathrm{SS}}=(1-\bar n)\,\delta_{mn}.
\end{equation}
Applying Wick's theorem to $\langle c_m^\dagger c_n c_j^\dagger c_j\rangle_{\mathrm{SS}}$ gives
\begin{align}
\langle c_m^\dagger c_n c_j^\dagger c_j\rangle_{\mathrm{SS}}
&=\langle c_m^\dagger c_n\rangle_{\mathrm{SS}}\langle c_j^\dagger c_j\rangle_{\mathrm{SS}}
+\langle c_m^\dagger c_j\rangle_{\mathrm{SS}}\langle c_n c_j^\dagger\rangle_{\mathrm{SS}}\nonumber\\
&=\bar n^2\,\delta_{mn}+\bar n(1-\bar n)\,\delta_{mj}\delta_{nj}.
\end{align}
Substituting into the expansion of $n_i(t)$ yields
\begin{align}
\langle n_i(t)n_j\rangle_{\mathrm{SS}}
&=\sum_{m,n}U_{im}^*(t)U_{in}(t)\langle c_m^\dagger c_n c_j^\dagger c_j\rangle_{\mathrm{SS}}\nonumber\\
&=\bar n^2\sum_m|U_{im}(t)|^2+\bar n(1-\bar n)\,|U_{ij}(t)|^2\nonumber\\
&=\bar n^2+\bar n(1-\bar n)\,|U_{ij}(t)|^2.
\label{sm:eq:nint_nj}
\end{align}
In the last line we used unitarity of the single-particle evolution, $\sum_m|U_{im}(t)|^2=1$.
Combining Eqs.~\eqref{sm:eq:mu2_unitary_basic}, \eqref{sm:eq:Nl2}, and \eqref{sm:eq:nint_nj}, one finds
\begin{align}
\langle N_\ell(t)N_\ell\rangle_{\mathrm{SS}}
&=\sum_{i,j\in\mathrm{seg}(\ell)}\langle n_i(t)n_j\rangle_{\mathrm{SS}}\nonumber\\
&=\ell^2\bar n^2+\bar n(1-\bar n)\sum_{i,j\in\mathrm{seg}(\ell)}|U_{ij}(t)|^2,
\end{align}
and therefore we arrive at the final expression for the second moment,
\begin{equation}
\mu_2^{(H)}(\ell,t)=2\bar n(1-\bar n)\left[\ell-\sum_{i,j\in\mathrm{seg}(\ell)}|U_{ij}(t)|^2\right].
\label{sm:eq:mu2_unitary_bessel_app}
\end{equation}
For a translation-invariant geometry,
$U_{ij}(t)=i^{\,j-i}J_{j-i}(Jt)$ and hence $|U_{ij}(t)|^2=J_{i-j}^2(Jt)$.
Writing the double sum in terms of relative coordinate $r=i-j$ gives
\begin{equation}
\sum_{i,j\in\mathrm{seg}(\ell)}|U_{ij}(t)|^2
=\sum_{r=-(\ell-1)}^{\ell-1}(\ell-|r|)\,J_r^2(Jt),
\end{equation}
which leads directly to Eq.~\eqref{sm:eq:mu2_unitary_bessel}.

\end{document}